\documentclass[twocolumn,10pt]{aastex631}

\usepackage{amsmath}
\usepackage{booktabs}

\begin{document}

\title{COOL-LAMPS. VII. Quantifying Strong-lens Scaling Relations with 177 Cluster-scale Strong Gravitational Lenses in DECaLS}
\shorttitle{Scaling Relations in Strong-lensing Galaxy Clusters}
\shortauthors{Mork et al.}

\correspondingauthor{Simon D. Mork}
\email{sdmork@asu.edu}
\suppressAffiliations

\author[0000-0002-5573-9131]{Simon D. Mork}
\affiliation{Department of Astronomy and Astrophysics, University of Chicago, 5640 S. Ellis Avenue, Chicago, IL 60637, USA}
\affiliation{School of Earth and Space Exploration, Arizona State University, 781 Terrace Mall, Tempe, AZ 85287, USA}
\affiliation{Beus Center for Cosmic Foundations, Arizona State University, 781 Terrace Mall, Tempe, AZ 85287, USA}

\author[0000-0003-1370-5010]{Michael D. Gladders}
\affiliation{Department of Astronomy and Astrophysics, University of Chicago, 5640 S. Ellis Avenue, Chicago, IL 60637, USA}
\affiliation{Kavli Institute for Cosmological Physics, University of Chicago, 5640 S. Ellis Avenue, Chicago, IL 60637, USA}

\author[0000-0002-3475-7648]{Gourav Khullar}
\affiliation{Department of Physics and Astronomy, University of Pittsburgh, 3941 O'Hara Street, Pittsburgh, PA 15260, USA}
\affiliation{Pittsburgh Particle Physics Astrophysics and Cosmology Center, University of Pittsburgh, 3941 O'Hara Street, Pittsburgh, PA 15260, USA}

\author[0000-0002-7559-0864]{Keren Sharon}
\affiliation{Department of Astronomy, University of Michigan, 1085 S. University Avenue, Ann Arbor, MI 48109, USA}

\author[0009-0005-1143-495X]{Nathalie Chicoine}
\affiliation{Department of Astronomy and Astrophysics, University of Chicago, 5640 S. Ellis Avenue, Chicago, IL 60637, USA}

\author[0000-0001-9978-2601]{Aidan P. Cloonan}
\affiliation{Department of Astronomy and Astrophysics, University of Chicago, 5640 S. Ellis Avenue, Chicago, IL 60637, USA}
\affiliation{Department of Astronomy, University of Massachusetts Amherst, 710 N. Pleasant Street, Amherst, MA 01003, USA}

\author[0000-0003-2200-5606]{H\r{a}kon Dahle} 
\affiliation{Institute of Theoretical Astrophysics, University of Oslo, P.O. Box 1029, Blindern, NO-0315 Oslo, Norway}

\author[0009-0003-0226-6988]{Diego Garza}
\affiliation{Department of Astronomy and Astrophysics, University of Chicago, 5640 S. Ellis Avenue, Chicago, IL 60637, USA}
\affiliation{Department of Astronomy and Astrophysics, University of California, Santa Cruz, CA 95064, USA}

\author[0000-0001-9816-0878]{Rowen Glusman}
\affiliation{Department of Astronomy and Astrophysics, University of Chicago, 5640 S. Ellis Avenue, Chicago, IL 60637, USA}
\affiliation{Gravitation \& Astroparticle Physics, University of Amsterdam, Science Park 904, 1098 XH Amsterdam, The Netherlands}

\author[0000-0003-2294-4187]{Katya Gozman}
\affiliation{Department of Astronomy and Astrophysics, University of Chicago, 5640 S. Ellis Avenue, Chicago, IL 60637, USA}
\affiliation{Department of Astronomy, University of Michigan, 1085 S. University Avenue, Ann Arbor, MI 48109, USA}

\author[0009-0006-6950-6351]{Gabriela Horwath}
\affiliation{Department of Astronomy and Astrophysics, University of Chicago, 5640 S. Ellis Avenue, Chicago, IL 60637, USA}

\author[0000-0001-8000-1959]{Benjamin C. Levine}
\affiliation{Department of Astronomy and Astrophysics, University of Chicago, 5640 S. Ellis Avenue, Chicago, IL 60637, USA}
\affiliation{Department of Physics and Astronomy, Stony Brook University, 100 Nicolls Road, Stony Brook, NY 11794, USA}

\author{Olina Liang}
\affiliation{Department of Astronomy and Astrophysics, University of Chicago, 5640 S. Ellis Avenue, Chicago, IL 60637, USA}

\author{Daniel Mahronic}
\affiliation{Department of Astronomy and Astrophysics, University of Chicago, 5640 S. Ellis Avenue, Chicago, IL 60637, USA}

\author[0000-0002-7113-0262]{Viraj Manwadkar}
\affiliation{Department of Astronomy and Astrophysics, University of Chicago, 5640 S. Ellis Avenue, Chicago, IL 60637, USA}
\affiliation{Department of Physics, Stanford University, 382 Via Pueblo, Stanford, CA 94305, USA}
\affiliation{Kavli Institute for Particle Astrophysics and Cosmology, Stanford University, 382 Via Pueblo, Stanford, CA 94305, USA}

\author[0000-0002-8397-8412]{Michael N. Martinez}
\affiliation{Department of Astronomy and Astrophysics, University of Chicago, 5640 S. Ellis Avenue, Chicago, IL 60637, USA}
\affiliation{Department of Physics, University of Wisconsin, Madison, 1150 University Avenue, Madison, WI 53706, USA}

\author[0000-0002-3361-2893]{Alexandra Masegian}
\affiliation{Department of Astronomy and Astrophysics, University of Chicago, 5640 S. Ellis Avenue, Chicago, IL 60637, USA}
\affiliation{Department of Astronomy, Columbia University, 538 W. 120th Street, New York, NY 10027, USA}

\author[0000-0001-9225-972X]{Owen S. Matthews Acu\~{n}a}
\affiliation{Department of Astronomy and Astrophysics, University of Chicago, 5640 S. Ellis Avenue, Chicago, IL 60637, USA}
\affiliation{Department of Astronomy, University of Wisconsin---Madison, 475 N. Charter Street, Madison, WI 53706, USA}

\author[0000-0001-5931-5056]{Kaiya Merz}
\affiliation{Department of Astronomy and Astrophysics, University of Chicago, 5640 S. Ellis Avenue, Chicago, IL 60637, USA}

\author[0000-0002-7922-9726]{Yue Pan}
\affiliation{Department of Astronomy and Astrophysics, University of Chicago, 5640 S. Ellis Avenue, Chicago, IL 60637, USA}
\affiliation{Department of Astrophysical Sciences, Princeton University, 4 Ivy Lane, Princeton, NJ 08544, USA}

\author[0000-0002-9142-6378]{Jorge A. Sanchez}
\affiliation{Department of Astronomy and Astrophysics, University of Chicago, 5640 S. Ellis Avenue, Chicago, IL 60637, USA}
\affiliation{School of Earth and Space Exploration, Arizona State University, 781 Terrace Mall, Tempe, AZ 85287, USA}

\author[0000-0002-2323-303X]{Isaac Sierra}
\affiliation{Department of Astronomy and Astrophysics, University of Chicago, 5640 S. Ellis Avenue, Chicago, IL 60637, USA}

\author[0000-0001-8008-7270]{Daniel J. Kavin Stein}
\affiliation{Department of Astronomy and Astrophysics, University of Chicago, 5640 S. Ellis Avenue, Chicago, IL 60637, USA}

\author[0000-0002-1106-4881]{Ezra Sukay}
\affiliation{Department of Astronomy and Astrophysics, University of Chicago, 5640 S. Ellis Avenue, Chicago, IL 60637, USA}
\affiliation{Department of Physics and Astronomy, Johns Hopkins University, 3400 N. Charles Street, Baltimore, MD 21218, USA}

\author[0009-0008-1518-8045]{Marcos Tamargo-Arizmendi}
\affiliation{Department of Astronomy and Astrophysics, University of Chicago, 5640 S. Ellis Avenue, Chicago, IL 60637, USA}

\author[0000-0001-6584-6144]{Kiyan Tavangar}
\affiliation{Department of Astronomy and Astrophysics, University of Chicago, 5640 S. Ellis Avenue, Chicago, IL 60637, USA}
\affiliation{Department of Astronomy, Columbia University, 538 W. 120th Street, New York, NY 10027, USA}

\author[0009-0002-9963-7564]{Ruoyang Tu}
\affiliation{Department of Astronomy and Astrophysics, University of Chicago, 5640 S. Ellis Avenue, Chicago, IL 60637, USA}
\affiliation{Department of Anthropology, Yale University, 10 Sachem Street, New Haven, CT 06520, USA}

\author[0000-0003-0295-875X]{Grace Wagner}
\affiliation{Department of Astronomy and Astrophysics, University of Chicago, 5640 S. Ellis Avenue, Chicago, IL 60637, USA}

\author[0000-0002-6779-4277]{Erik A. Zaborowski}
\affiliation{Department of Astronomy and Astrophysics, University of Chicago, 5640 S. Ellis Avenue, Chicago, IL 60637, USA}
\affiliation{Department of Physics, The Ohio State University, 191 W. Woodruff Avenue, Columbus, OH 43210, USA}
\affiliation{Center for Cosmology and Astro-Particle Physics, The Ohio State University, 191 W. Woodruff Avenue, Columbus, OH 43210, USA}

\author[0000-0001-6454-1699]{Yunchong Zhang}
\affiliation{Department of Astronomy and Astrophysics, University of Chicago, 5640 S. Ellis Avenue, Chicago, IL 60637, USA}
\affiliation{Department of Physics and Astronomy, University of Pittsburgh, 3941 O'Hara Street, Pittsburgh, PA 15260, USA}

\collaboration{30}{(COOL-LAMPS Collaboration)}

\received{2024 January 12}
\revised{2024 November 29}
\accepted{2024 December 8}
\published{2025 January 27}
\submitjournal{\apj}

\begin{abstract}
We estimate the Einstein-radius-enclosed total mass for 177 cluster-scale strong gravitational lenses identified by the ChicagO Optically selected Lenses Located At the Margins of Public Surveys (COOL-LAMPS) collaboration with lens redshifts ranging from $0.2 \lessapprox z \lessapprox 1.0$ using the brightest-cluster-galaxy (BCG) redshift and an observable proxy for the Einstein radius. We constrain the Einstein-radius-enclosed luminosity and stellar mass by fitting parametric spectral energy distributions to aperture photometry from the Dark Energy Camera Legacy Survey in the $g$-, $r$-, and $z$-band Dark Energy Camera filters. We find that the BCG redshift, enclosed total mass, and enclosed luminosity are strongly correlated and well described by a planar relationship in 3D space. We find that the enclosed total mass and stellar mass are correlated with a logarithmic slope of $0.500^{+0.029}_{-0.031}$, and the enclosed total mass and stellar-to-total mass fraction are correlated with a logarithmic slope of $-0.495^{+0.032}_{-0.033}$. In tandem with the small radii within which these slopes are constrained, this may suggest invariance in baryon conversion efficiency and feedback strength as a function of cluster-centric radii in galaxy clusters. Additionally, the correlations described here should have utility in ranking strong-lensing candidates in upcoming imaging surveys---such as Rubin/Legacy Survey of Space and Time---in which an algorithmic treatment of strong lenses will be needed due to the sheer volume of data these surveys will produce.
\end{abstract}

\keywords{Galaxy clusters (\href{http://astrothesaurus.org/uat/584}{584}) ; High-redshift galaxy clusters (\href{http://astrothesaurus.org/uat/2007}{2007}) ; Scaling relations (\href{http://astrothesaurus.org/uat/2031}{2031}) ; Spectral energy distribution (\href{http://astrothesaurus.org/uat/2129}{2129}) ; Strong gravitational lensing (\href{http://astrothesaurus.org/uat/1643}{1643})}

\section{Introduction}

\begin{figure}[!ht]
    \centering
    \includegraphics[width=8.5cm]{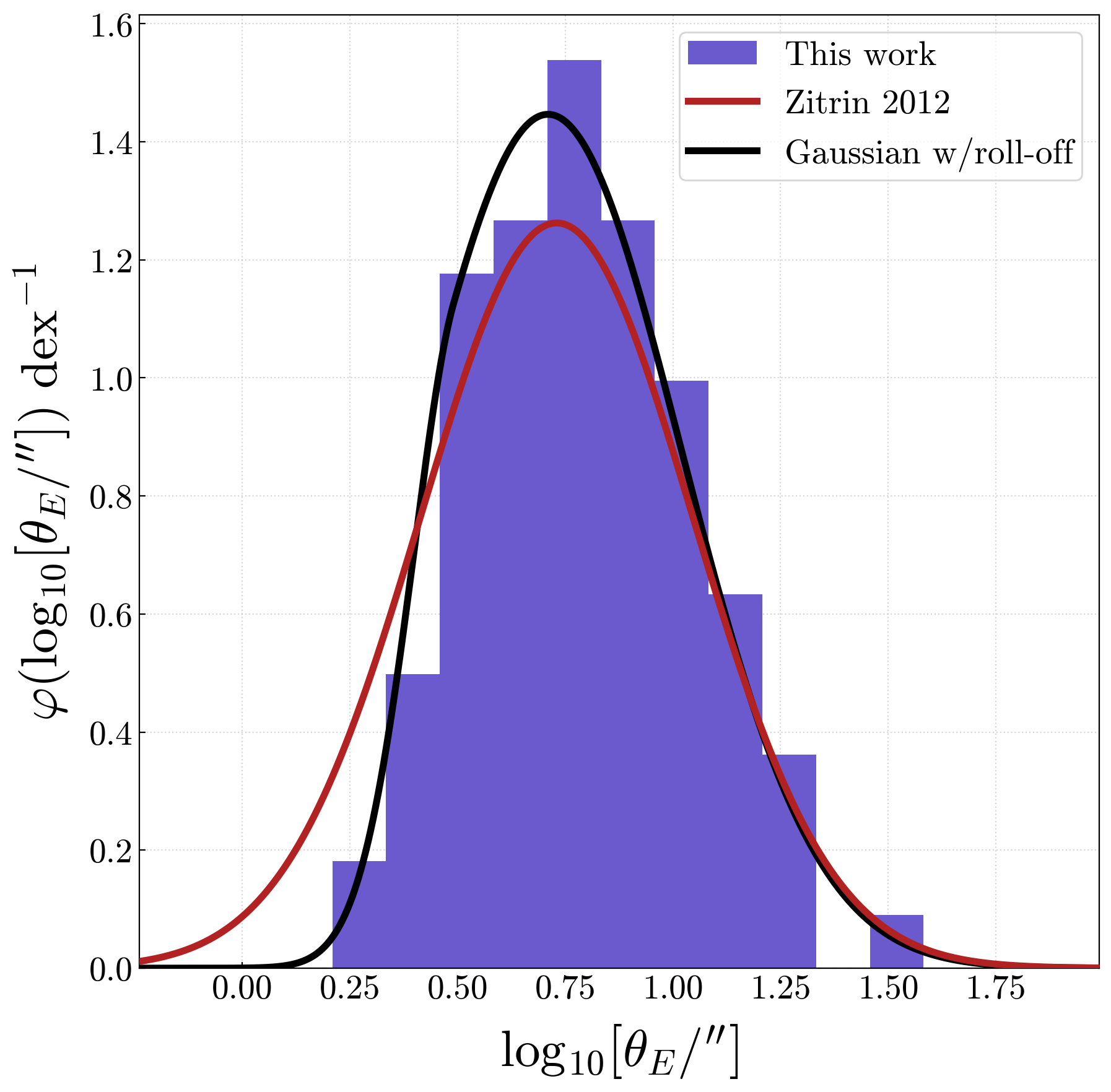}
    \caption{A normalized histogram of the fitted Einstein radii in log space for this work shown in blue. Overplotted in black is a normalized Gaussian with one-sided Gaussian roll-off fit to the cumulative distribution function (CDF) of these radii in dex (see Section \ref{Einstein Radius} for parameters describing this fit). Also plotted in red is a Gaussian probability distribution function with $\mu = 0.73$ and $\sigma = 0.316$ from \cite{2012MNRAS.423.2308Z} representing the simulated Einstein radius distribution of 10,000 Sloan Digital Sky Survey (SDSS) clusters.}
    \label{einstein_zitrin}
\end{figure}

Strong gravitational lensing is a rarely observed phenomenon in the Universe in which an intervening object with sufficient surface-mass density bends light from a background source relative to the observer such that multiple images of the source are formed. For cases where the background source is a galaxy, the lensing effect can create highly distorted and magnified images of the source galaxy in an arc-like shape (e.g., \citealt{1986BAAS...18R1014L, 1987A&A...172L..14S, 2011A&ARv..19...47K, 2013SSRv..177...31M, 2017A&A...608L...4R, 2024SSRv..220...87S}). The discovery of new strong lenses is accelerating rapidly, particularly with the help of machine learning applied to large optical-imaging surveys (e.g., \citealt{2021ApJ...909...27H,2022A&A...668A..73R, 2023ApJ...954...68Z}). Coupled with access to a flood of new data from large surveys such as Rubin/Legacy Survey of Space and Time (LSST) in the coming years, the population of candidate lenses both in the bulk and at the margins will only grow. 

Any flux-limited survey will have tentative arcs; i.e. at the flux limit. Regardless of the lensing configuration, numerous faint features that are suggestive of strong lensing, but whose reality and interpretation as strong lensing, will exist and require verification. Contemporary methodologies of identifying strong lenses often rely on visual inspection in order to winnow candidate lists to an acceptable number worthy of detailed follow-up (e.g., \citealt{2017ApJS..232...15D}); such inspection has been successful in locating new systems thus far (e.g., \citealt{2023MNRAS.523.4413R}). In addition, automated strong-lens modeling has emerged as a promising avenue toward blindly identifying multiple-image families in strong lenses (e.g., \citealt{2019MNRAS.482.1824S, 2020MNRAS.491.3778C, 2020MNRAS.498.1121Z}), though such methodologies typically rely on a light-traces-mass model to feed to a preliminary lens model---of which the latter can be computationally expensive. With thousands to millions of lines of sight to choose from in an ever-growing plethora of imaging data, analyzing statistical measures and correlations in existing strong lenses may help hone candidate-lens samples by quantifying instances of cluster-scale strong lensing in tandem with geometric identification. 

\begin{figure*}[!ht]
    \centering
    \includegraphics[width=18cm]{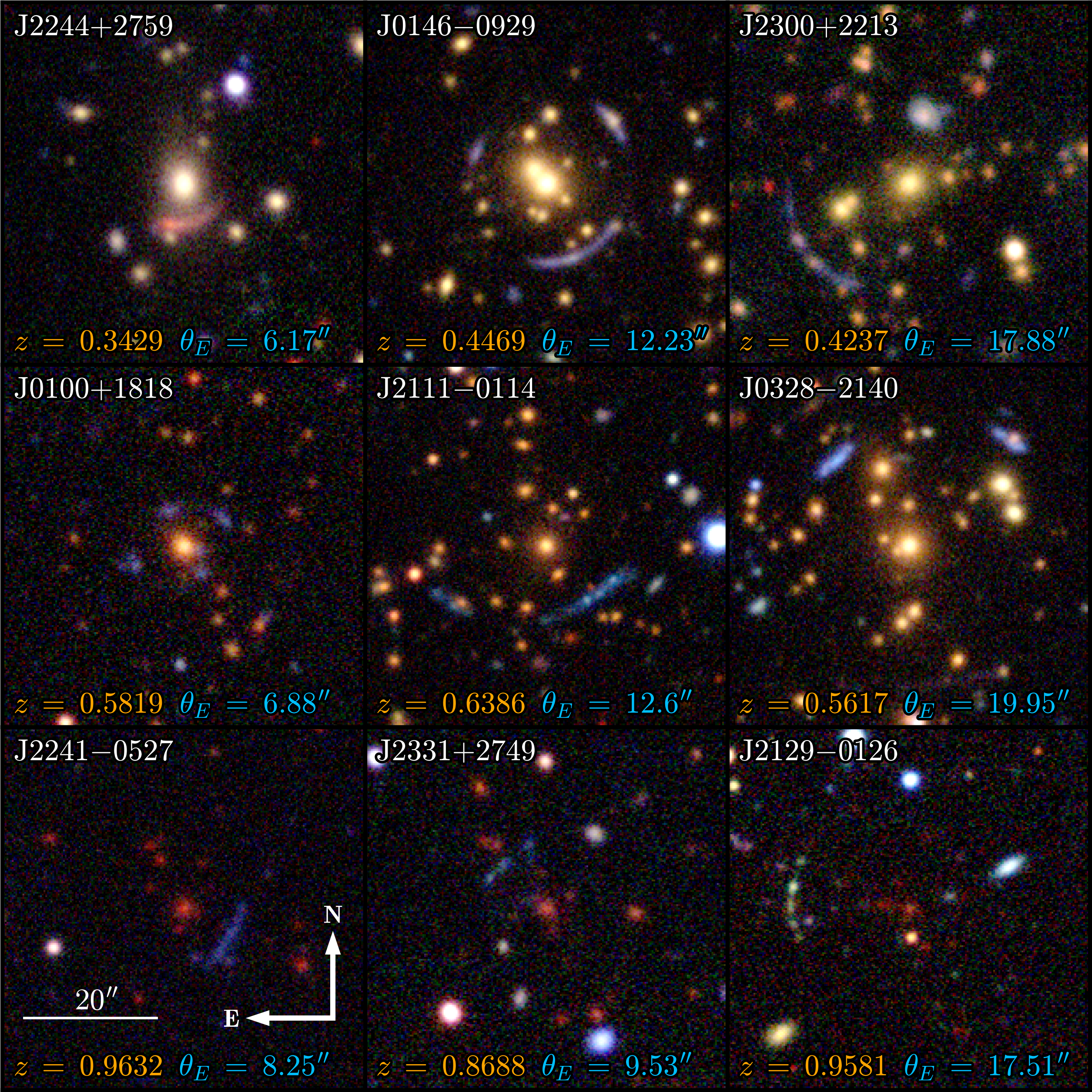}
    \caption{A subset of nine strong gravitational lenses analyzed in this work. The lenses are arranged by increasing redshift from the top to the bottom of the Figure, and they are arranged by increasing Einstein radius from the left to the right of the Figure. Respective lens redshifts and Einstein radii are also shown in each subplot. RGB images have equal astrometric scaling and are sourced from DECaLS LS DR9 \citep{2019AJ....157..168D} using $z$-, $r$-, and $g$-band imaging data with custom color scaling, respectively. This Figure is available as an animation. The animation fades in and out red circles, which denote positional constraints used to derive the Einstein radius, white circles representing the Einstein aperture, and a black `x' marking the exact coordinates of each BCG. (An animation of this figure is available in the \href{https://doi.org/10.3847/1538-4357/ada24c}{online article}.)}
    \label{lens_collage}
\end{figure*}

This work aims to quantify scaling relations in existing cluster-scale strong lenses such that ambiguous systems could be analyzed with a similar methodology to determine if a given system lies on or off the scaling relationships we derive here. Coupled with existing algorithms that can identify preferential lines of sight for cluster-scale strong lensing (e.g., \citealt{2013ApJ...769...52W}) and an arc-finding algorithm (e.g., \citealt{2007A&A...472..341S}), these calculations could be undertaken automatically for arbitrary imaging data. This analysis would help streamline the ability to identify targets for follow-up observations by ranking candidate lenses to produce a purer sample. We analyze a sample of 177 cluster-scale strong gravitational lenses identified via visual inspection by the COOL-LAMPS collaboration in DECaLS Legacy Survey Data Release 8 (LS DR8, \citealt{2019AJ....157..168D}) images, and we investigate the scaling relations between the brightest-cluster-galaxy (BCG) redshift, Einstein-radius-enclosed (enclosed) total mass, enclosed luminosity, enclosed stellar mass, and enclosed stellar-to-total mass fraction therein. The sample we use here consists of a refined subsample of all strong-lensing candidates identified by COOL-LAMPS that we consider robust---despite a lack of definite spectroscopic confirmation in many cases. All of the lenses used here have the visual geometry of a giant arc with clear curvature that is bright enough to be unambiguously visible in the survey data.

\begin{figure*}[!ht]
    \centering
    \includegraphics[width=18cm]{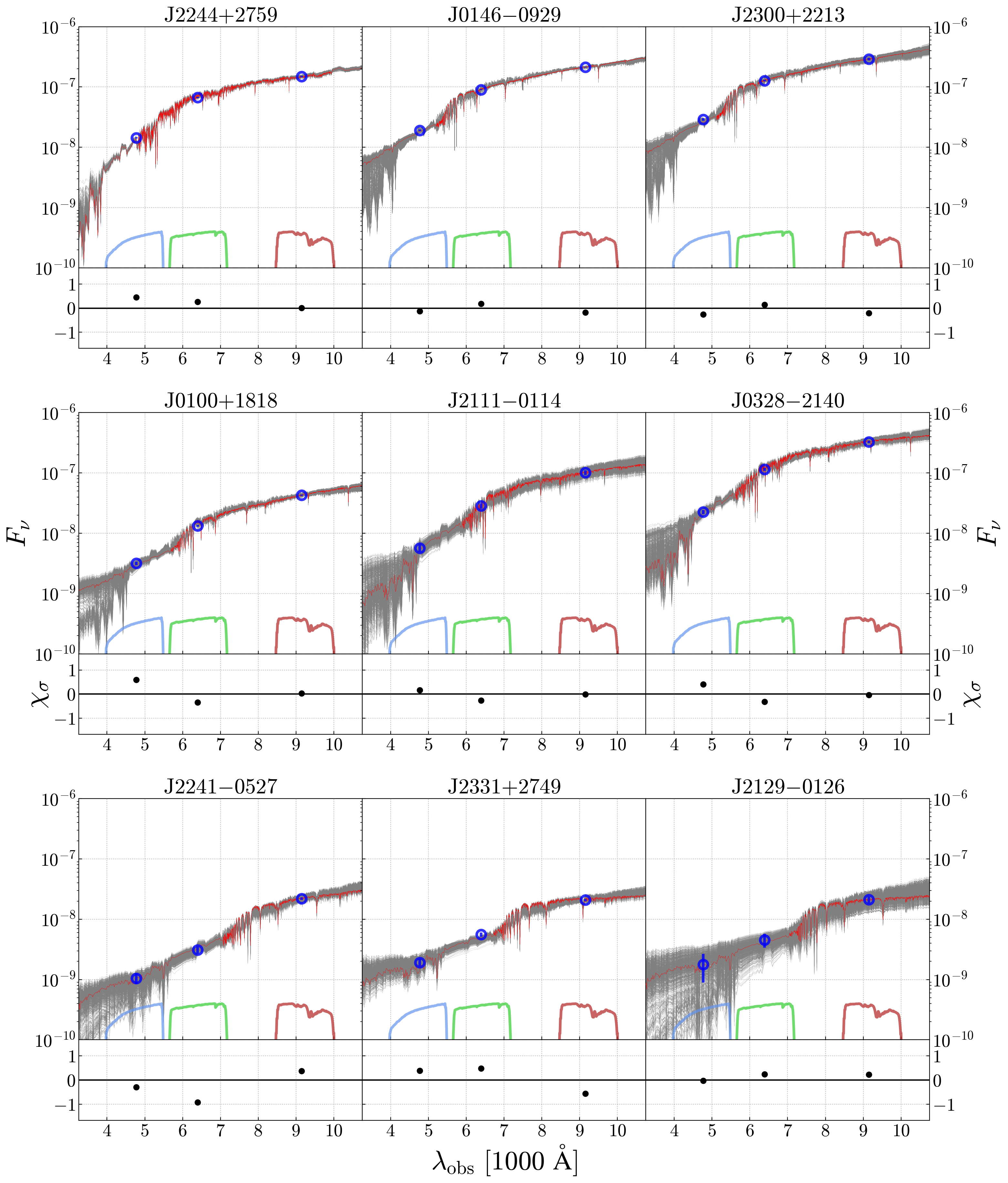}
    \caption{Observed-frame spectral energy distributions (SEDs) in maggies for the nine lenses shown in Figure \ref{lens_collage} after SED fitting with \texttt{Prospector}. The top plot of each subplot contains the best-fit SED (red line), 1024 SEDs drawn from their respective posterior distributions (gray lines), observed cluster photometry (blue points), and photometric filter transmission curves for the Dark Energy Camera (DECam; \citealt{2015AJ....150..150F}) $g$-, $r$-, and $z$-band filters from left to right, respectively (in arbitrary units). The bottom plot of each subplot shows the residual between the observed photometry and the best-fit SED---normalized by the standard deviation of the observed photometry.}
    \label{onespectrum}
\end{figure*}

Throughout this paper, we adopt a flat cold dark matter WMAP-9 cosmology \citep{2013ApJS..208...19H}. All photometric calculations were done in the AB magnitude system.

\section{Methodology}
Galaxies and galaxy clusters host their mass in a variety of forms. Theory and observations both place strong constraints on observed stellar luminosity as a function of mass (e.g., \citealt{1938ApJ....88..472K,2014A&A...565A.126P,2018A&A...619L...1W}), hot intracluster gas and diffuse intracluster light contribute to the total mass in galaxy clusters (e.g., \citealt{1972ApJ...178..309F, 1972ApJ...174L..65K, 2005ApJ...618..195G, 2007PhR...443....1M}), and both galaxies and galaxy clusters are composed of a significant fraction of dark matter (e.g., \citealt{1933AcHPh...6..110Z, 1986HiA.....7...27R,1996ApJ...462..563N, 1998ApJ...495...80B, 2000ApJ...543..521H}). However, robustly measuring these mass components on a per-system basis is time and resource intensive---which is unsustainable for large samples. Conveniently, the existence of lensing prescribes a specific amount of mass that must exist to create the observed lensing \textit{a priori}.

\cite{2020ApJ...902...44R} have shown that the core mass interior to the Einstein radius of a strong-lensing galaxy cluster may be computed with an uncertainty of $\approx10\%$ via the simple evaluation of a function parameterized by the angle between the BCG and lensed source arc (a proxy for Einstein radius; $\theta_E$) and the line-of-sight geometry (redshift of the lens and the source; $z_L$, $z_S$). With these three parameters, the mass can be immediately calculated and accounts for \textit{all} of the projected mass density interior to the Einstein radius for a strong lens whose dark matter halo can be described by a circularly symmetric dark matter profile. Measuring mass in this way also offers a well-defined aperture with which to constrain photometric properties of the strong lens as well. By directly integrating over pixels that correspond to cluster members inside this aperture in flux-calibrated images, the photometry of the strong lens can be automatically utilized by a spectral energy distribution (SED) modeler to obtain measurements of enclosed luminosity and stellar mass.

The COOL-LAMPS collaboration has constructed and analyzed a large dataset of strong-lensing candidates using a simple human-led ranking system (e.g., \citealt{2021ApJ...906..107K, 2022ApJ...940...42S, 2023ApJ...946...63M, 2023ApJ...950...58Z, 2023ApJ...954L..38N, 2024ApJ...963...44K, 2024arXiv240803379C}). From this, we have collated a sample of 177 galaxy clusters that exhibit robust visual evidence of strong gravitational lensing in which the primary lensed source arc is roughly circular with its projected center located at or very near to the BCG. Any systems with complex source-galaxy geometry in the image plane imply that the lens possesses a complex dark matter distribution that is less likely to be well described a circularly symmetric dark matter profile. Since the assumption of a circularly symmetric dark matter profile underpins the functions used to estimate the enclosed mass described in \cite{2020ApJ...902...44R}, such complex systems were intentionally excluded from this work.

\subsection{Einstein Radius} \label{Einstein Radius}
We defined the Einstein radius for each system as the radius of a circle centered on the BCG \citep{2020ApJ...902...44R} that minimizes the total angular separation between each of the three most readily identifiable bright ``clumps” in the tangentially lensed arcs and the perimeter of said circle. Henceforth, we refer to this circular aperture centered on the BCG with radius equal to the Einstein radius as the ``Einstein aperture". Four points (the BCG center and three points along each arc) physically constrain the Einstein aperture, and we obtained them by tagging the point of peak surface brightness for the BCG and similar peaks in surface brightness in the clumps of each tangentially lensed arc, respectively. If three distinct peaks were unable to be found along the tangential arcs, the main arc was simply traced. Each tangential-arc point was assigned a Gaussian uncertainty in both spatial axes with $\mu$ equal to its location and $\sigma = 0.5^{\prime\prime}$ which roughly corresponds to the width of an average arc. We bootstrapped the three tangential-arc points 1000 times, with each sampling drawn from the 2D Gaussian distribution for each point. We adopted the mean and standard deviation from the resulting distribution of Einstein radii as the Einstein radius and its error, respectively, for each system. By fitting to the cumulative distribution function (CDF) of the Einstein radii for all 177 systems in this work, we find that they are distributed log normally with $\mu=0.71\pm0.07$ and $\sigma=0.31\pm0.04$. We also parameterize an incompleteness roll-off to low values with a one-sided Gaussian centered at $0.49\pm0.16$ with $\sigma=0.14\pm0.07$ which indicates that our sample is incomplete starting at $3.09^{\prime\prime}\,^{+1.37}_{-0.99}$ and 50\% incomplete by $2.11^{\prime\prime}\,^{+1.08}_{-0.74}$. Physical motivations for this incompleteness are further discussed in Section \ref{photo_err}. \cite{2012MNRAS.423.2308Z} simulated the ``universal'' distribution of Einstein radii for a sample of 10,000 Sloan Digital Sky Survey (SDSS) clusters and found that this distribution is log normal with $\mu=0.73^{+0.02}_{-0.03}$ and $\sigma=0.316^{+0.004}_{-0.002}$. As shown in Figure \ref{einstein_zitrin}, we find good agreement between the distribution of Einstein radii in this work and that of \cite{2012MNRAS.423.2308Z}---indicating that our methodology is a good proxy for the true underlying Einstein radius for our sample of galaxy clusters.

\subsection{Lens Redshift}
BCGs are \textit{sui generis} red-sequence cluster galaxies that can also be used as proxies to infer properties of the cluster-scale dark matter halo (e.g., \citealt{1991ApJ...375...15O, 2014ApJ...797...82L}). Since we assume that the strong lenses in this work can be well described by a symmetric dark matter profile, we take the redshift of the BCG as interchangeable with the systematic redshift of the cluster-scale dark matter halo. The lens redshift converts the measured Einstein radius to a proper transverse distance via the angular diameter distance, and we queried the SDSS Data Release 15 (DR15, \citealt{2019ApJS..240...23A, 2019BAAS...51g.274K}) in tandem with the Legacy Survey Data Release 9 (LS DR9, \citealt{2019AJ....157..168D, 2022MNRAS.512.3662D}) for each system in this work to obtain lens (BCG) redshifts. If the centroid coordinates for each BCG corresponded to a warning-free spectroscopic redshift in SDSS DR15, we adopted that redshift as the lens redshift for that system along with its associated error. If a given BCG lacked any corresponding value in SDSS DR15, as is the case for the higher-redshift clusters in this work and/or clusters located in SDSS-non-imaged areas, we adopted the LS DR9 photometric redshift \citep{2023JCAP...11..097Z} as the lens redshift and its associated error instead. We obtained spectroscopic redshifts for 41 systems and photometric redshifts for the remaining 136 systems. A descriptive visualization sampling the range of Einstein radii and lens redshifts considered in this work is shown in Figure \ref{lens_collage}.

\begin{deluxetable*}{clll}
\tabletypesize{\small}
\tablecolumns{4}
\tablecaption{Parameters Used in SED Fitting with \texttt{Prospector} \label{tab:prosparams}}
\tablehead{\colhead{Free} & \colhead{Parameter} & \colhead{Description} & \colhead{Priors}}
\startdata 
Y & $\mathrm{log}(M_{tot}/M_{\odot})$ & Total stellar mass formed in dex solar masses. & Top Hat: [8.0, 14.0]. \\
Y & $\mathrm{log}(Z/Z_{\odot})$ & Stellar metallicity in dex solar metallicity. & Top Hat: [-1.0, 0.2]. \\
Y & $\lambda_2$ & Diffuse dust optical depth. & Top Hat: [0.0, 2.0]. \\
Y & $t_{\mathrm{age}}$ & Age of the cluster in gigayears. & Top Hat: [$t_{z=20}$, $t_{z_{BCG}}$]. \\
Y & $\tau$ & SFH e-folding time in gigayears. & Top Hat: [0.1, 10]. \\
N & imf\_type & Initial mass function type. & Chabrier \citep{2003PASP..115..763C}. \\
N & dust\_type & Dust attenuation curve. & Calzetti \citep{2000ApJ...533..682C}. \\
N & sfh & Star formation history model. & Delayed tau \citep{2019ApJ...873...44C}. \\
\enddata
\end{deluxetable*}

\subsection{Source Redshift}
In strong lensing, the source redshift is typically constrained either with spectroscopic follow-up (e.g., \citealt{2020ApJS..247...12S}) or by inferring a photometric redshift (e.g., \citealt{2018ApJ...859..159C}). However, obtaining these redshifts for all identified source galaxies in a given system requires prolonged additional inquiry in tension with the appeal of an efficient mass estimator. \cite{2020ApJ...902...44R} showed that substituting a single known source-galaxy redshift with a distribution of source-galaxy redshifts introduces a statistically insignificant uncertainty into the final distribution of mass measurements when compared to the magnitude of other systematic uncertainties. This distribution of lensed source redshifts is measurable (e.g., \citealt{2011ApJ...727L..26B, 2011ApJS..193....8B, 2022AJ....164..148T}), and we simply adopt a well-described Gaussian with $\mu=2$ and $\sigma=0.2$ from \cite{2011ApJ...727L..26B} in keeping with the methodology of \cite{2020ApJ...902...44R} as the distribution of source redshifts across the entire sample in this work. We need not consider the redshift distribution of sources other than giant arcs, since all of the source galaxies we analyze in this work are geometrically consistent with the giant arcs analyzed in \cite{2011ApJ...727L..26B}. While we note that a later paper by \cite{2012ApJ...744..156B} favors a slightly smaller error with the same mean, using this distribution would only marginally impact the uncertainties on the calculated total masses. We therefore use the distribution in \cite{2011ApJ...727L..26B} to maintain continuity with \cite{2020ApJ...902...44R}.


\subsection{Photometry} \label{sec_photometry}
We obtained aperture photometry for SED fitting in each system by linearly summing the dereddened flux from pixels within the Einstein aperture in $g$-, $r$-, and $z$-band imaging data from DECaLS LS DR9. Since it is critical that we do not include light from the lensed source arcs as well as any interloping stars or galaxies that are not a member of the lensing cluster in the measured lens photometry, we created two masks to exclude non-cluster-member regions. The first mask removed flux from the lensed source arcs, and the second mask removed flux from interlopers as identified by a color and magnitude screening by eye. Accounting for masking, this linear integration was done for each of the 1000 bootstrapped Einstein apertures derived in Section \ref{Einstein Radius}. The resulting mean was adopted as the $g$-, $r$-, and $z$-band photometry for each system---with the standard deviation quantifying the geometric error. In quantifying the photometric error per-band, we first fit a sigma-clipped Gaussian to a histogram of the pixel values characterizing the background in each band for each system. The standard deviation of this Gaussian multiplied by $\sqrt{N}$, where $N$ is the number of pixels interior to the Einstein aperture, quantifies the statistical uncertainty in the photometric error for each band in each system. Next, the standard deviations for each Gaussian in a given photometric band across all systems were plotted as a histogram and fitted with another Gaussian. The standard deviation of this second Gaussian multiplied by $N$ characterizes the systematic uncertainty of the sky subtraction for each band across all systems. Taken together, the quadrature sum of the geometric Einstein aperture error, statistical photometric error, and systematic sky-subtraction error were adopted as the $g$-, $r$-, and $z$-band error for each system.

\section{Analysis} \label{sec_analysis}

\subsection{Total Mass}
Following \cite{2020ApJ...902...44R}, the enclosed total mass within the Einstein aperture was calculated using Equations \eqref{mass_eq_1} and \eqref{mass_eq_2}---where $D(z)$ refers to the angular diameter distance at redshift $z$---via a Monte Carlo approach. The enclosed total mass was computed 1000 times for each system, in which each calculation used a different Einstein radius, source redshift, and lens redshift randomly drawn from a Gaussian distribution with mean equal to the parameter value and standard deviation equal to the parameter error. Also described in \cite{2020ApJ...902...44R} is the need to apply an empirical correction described in Equation \eqref{mass_eq_3} based on the ``completeness" of the lensed source arcs in each system---where $f(\theta_E)$ is a cubic polynomial function specified in Table 1 of \cite{2020ApJ...902...44R}. This is because the systematic bias and scatter of $M_{\Sigma}(<\theta_E)$ has a dependence on the degree of symmetry for each system---in which larger-Einstein-radius systems have a stronger offset. The correction factor aims to remove this dependence as a function of Einstein radius in non-perfectly symmetric systems. 

\begin{equation}
    \Sigma_{\mathrm{cr}}(z_L,z_S) = \frac{c^2}{4\pi G}\frac{D_S(z_S)}{D_L(z_L)D_{LS}(z_L,z_S)}
    \label{mass_eq_1}
\end{equation}
\begin{equation}
    M(<\theta_E) = \pi(D_L(z_L)\theta_E)^2\,\Sigma_{\mathrm{cr}}(z_L,z_S)
    \label{mass_eq_2}
\end{equation}
\begin{equation}
    \mathrm\ M_{\Sigma}(<\theta_E) = \frac{M(<\theta_E)}{f(\theta_E)}
    \label{mass_eq_3}
\end{equation}

For the purposes of applying this correction in each of the 177 strong lenses analyzed in this work, we measured the azimuthal coverage ($\phi$)---defined as the percentage of the Einstein aperture that was traced out by the lensed source arcs in a given system. For example, a lens with a tangential arc stretching from an azimuthal angle of $10^{\circ}$ to $100^{\circ}$ would have $\phi=0.25$. As was done by \cite{2020ApJ...902...44R}, $\phi$ was used as an observable proxy for the degree of symmetry in each system to determine whether or not the correction factor was needed. Larger-Einstein-radius systems with a predominantly low $\phi$ tend to deviate from a symmetric dark matter halo and thus require the empirical correction. For spherically symmetric systems with a large $\phi$, the measured $M_{\Sigma}(<\theta_E)$ is taken to be fairly unbiased, and thus the correction factor is not needed. We obtained $\phi$ by determining the fraction of the perimeter of the Einstein aperture that was subtended by the regions masking the lensed source arcs described in Section \ref{sec_photometry} in each system. Adopting the convention used by \cite{2020ApJ...902...44R} in their analysis, if $\phi<0.5$, Equation \eqref{mass_eq_3} was applied. If $\phi\ge0.5$, it was not applied. Accounting for the empirical correction, the mean and standard deviation of the 1000 measurements of $M_{\Sigma}(<\theta_E)$ were adopted as the enclosed total mass and its error, respectively, for each system.

\subsection{Stellar Luminosity and Mass}
We conducted parametric SED fitting using the photometry obtained in Section \ref{sec_photometry} with \texttt{Prospector} \citep{2009ApJ...699..486C, 2010ApJ...712..833C, 2021ApJS..254...22J, ben_johnson_2023_10026684}; five free parameters and three assumed parameters parameterized each SED. A summary of all eight parameters used in fitting can be found in Table \ref{tab:prosparams}. While the number of free parameters used in fitting is greater than the number of photometric band constraints for each system, parameters such as age, metallicity, and dust are correlated with one another (e.g., \citealt{1994ApJS...95..107W, 2017A&A...602A..35T}), and thus, SED fitting as done here remains sensible. Utilizing \texttt{emcee} \citep{2013PASP..125..306F} as implemented in \texttt{Prospector} with 84 walkers for a total of 6720 iterations, only the last 840 iterations for each of the 84 walkers (70,560 total parameter vectors) were taken as representing the posterior distribution for each of the five free parameters in each system in order to eliminate the significant burn-in sequence of the fitting process. After fitting, we adopted the 50th percentile of the posterior distribution as the value for each free parameter, and the greater difference between the 84th-50th percentile and 50th-16th percentile values as the error. Observed-frame SEDs for the systems shown in Figure \ref{lens_collage} after SED fitting are shown in Figure \ref{onespectrum}.

\begin{figure*}[!ht]
    \centering
    \includegraphics[width=18cm]{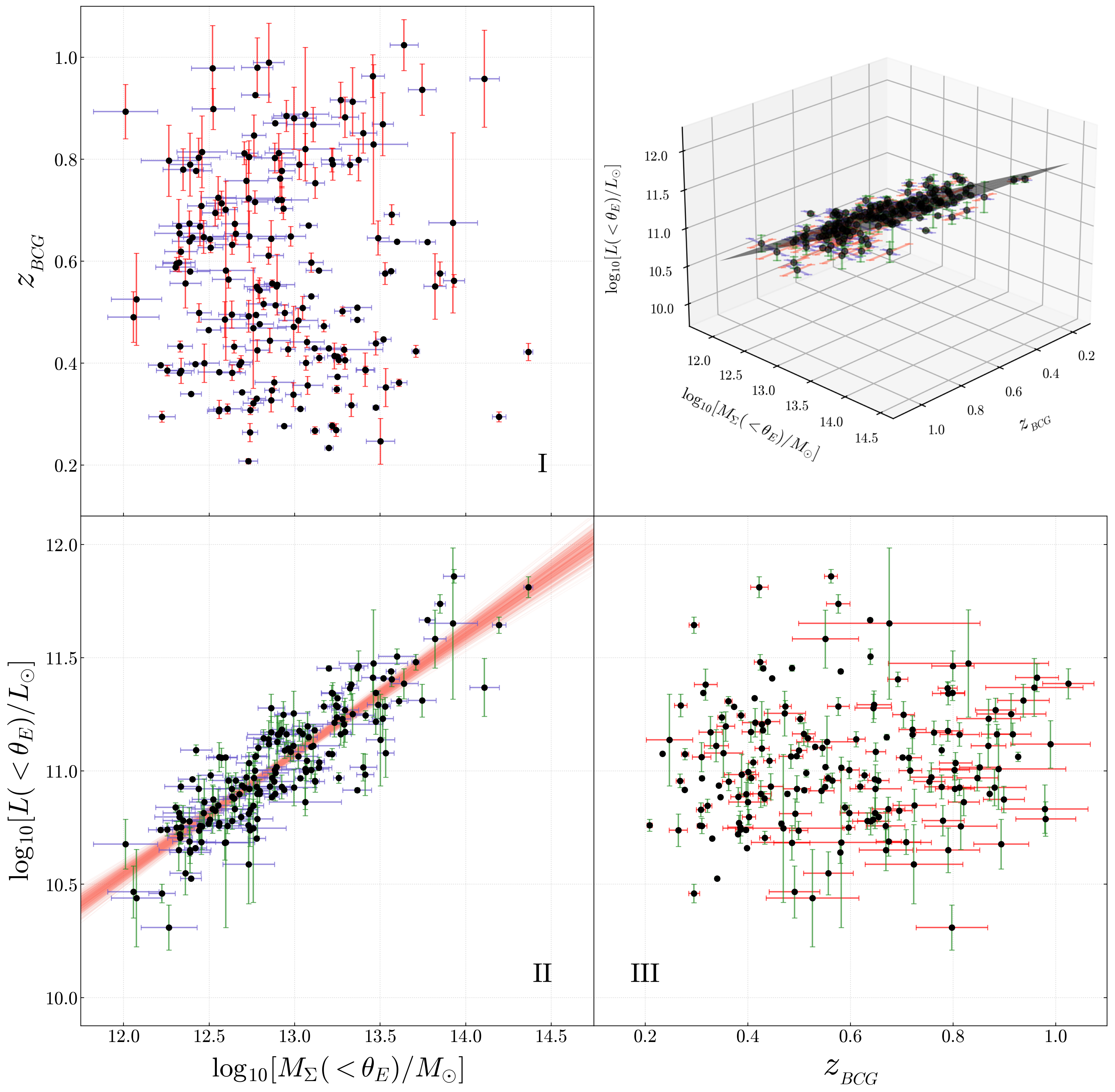}
    \caption{A corner plot of the BCG redshift, enclosed total mass, and enclosed luminosity for the systems in this work plotted as black circles. Errors for each variable in each subplot are represented as red, blue, and green lines, respectively. A plane-of-best-fit is also shown in the top-right plot as well. This Figure is available as an animation. The animation depicts a rotating view of the plane and its associated constraints shown in the top right of the static Figure to emphasize its features in 3D. (An animation of this figure is available in the \href{https://doi.org/10.3847/1538-4357/ada24c}{online article}.)}
    \label{3d_plane_plot}
\end{figure*}

To derive luminosity, we generated new SEDs with a random sample of 1024 free-parameter vectors from the posterior distributions of the best-fit SED in each system. We integrated over the rest-frame wavelength interval of 3000--7000\r{A}, since this corresponds to the longest wavelength interval sampled by part of at least two bands of photometry across all BCG redshifts in this work (the lower bound of 3000\r{A} gets redshifted out of the $\mathit{g}$ and into the $\mathit{r}$ band at high redshifts). This range also allows us to sample the rest-frame 4000\r{A} break while recovering flux in longer wavelengths, which are significantly more luminous in cluster galaxies. After integrating the SEDs, we converted flux to luminosity in solar luminosities using the appropriate luminosity distance and adopted the mean and standard deviation from the 1024 random-parameter vectors as the luminosity and its error, respectively, for each system.

\section{Results}

As Figure \ref{3d_plane_plot} shows, the BCG redshift, enclosed total mass, and enclosed luminosity are correlated. By performing a linear regression scatter using \texttt{emcee} that accounts for errors in data on both axes \citep{2010arXiv1008.4686H}, we find the slope of the logarithmic correlation between the enclosed total mass and enclosed luminosity to be $0.528^{+0.022}_{-0.023}$ with an intrinsic scatter in luminosity of $0.115^{+0.008}_{-0.007}$ dex from the posterior distribution of the Markov Chain Monte Carlo (MCMC) fit. Indeed, we note that the most-massive clusters in this work also possess a large stellar mass and thus have a large luminosity. Also using \texttt{emcee}, we constrained a plane-of-best-fit taking into account the error in data on all three axes to describe this correlation---where X = $z_{_{BCG}}$, Y = $\mathrm{log}_{10}[M_{\Sigma} (<\theta_E) / M_{\odot}]$, and Z = $\mathrm{log}_{10}[L(<\theta_{E})/L_\odot]$ with an intrinsic scatter on Z of $0.110^{+0.009}_{-0.008}$ dex---yielding:

\begin{equation}
    Z = 0.148^{+0.051}_{-0.052}X + 0.529^{+0.023}_{-0.022}Y + 4.121^{+0.272}_{-0.298}
    \label{plane_eq}
\end{equation}

The strong lensing candidates used to derive this relationship are, in our opinion, robust and unlikely to be non-lensing objects misclassified as \textit{bona fide} lenses. Within the COOL-LAMPS candidate sample from which this subset derives, and indeed any other sample of strong lens candidates, there are a much larger number of less robust candidate systems. The identification of these systems as strong lenses may be more uncertain due to less obvious lensing geometry, and also---as is true in all flux-limited surveys---many of the notional lensed images will be only ambiguously detected. However, the simple and analytic relationship above, derived from a subsample of robust candidate strong lenses considering only quantities derived from the discovery survey observations, may have utility in differentiating \textit{bona fide} strong lenses from non-lenses in the correspondingly larger samples of candidates where the evidence of lensing is not nearly as robust as the sample of candidates analyzed here. Moreover, this correlation offers a future opportunity to test simulations against observations---particularly those that aim to generate realistic galaxy populations in dense cluster environments.

A full accounting of all measured quantities in this work can be found in Table \ref{tab:sample}. 

\begin{figure*}[!ht]
    \centering
    \includegraphics[width=18cm]{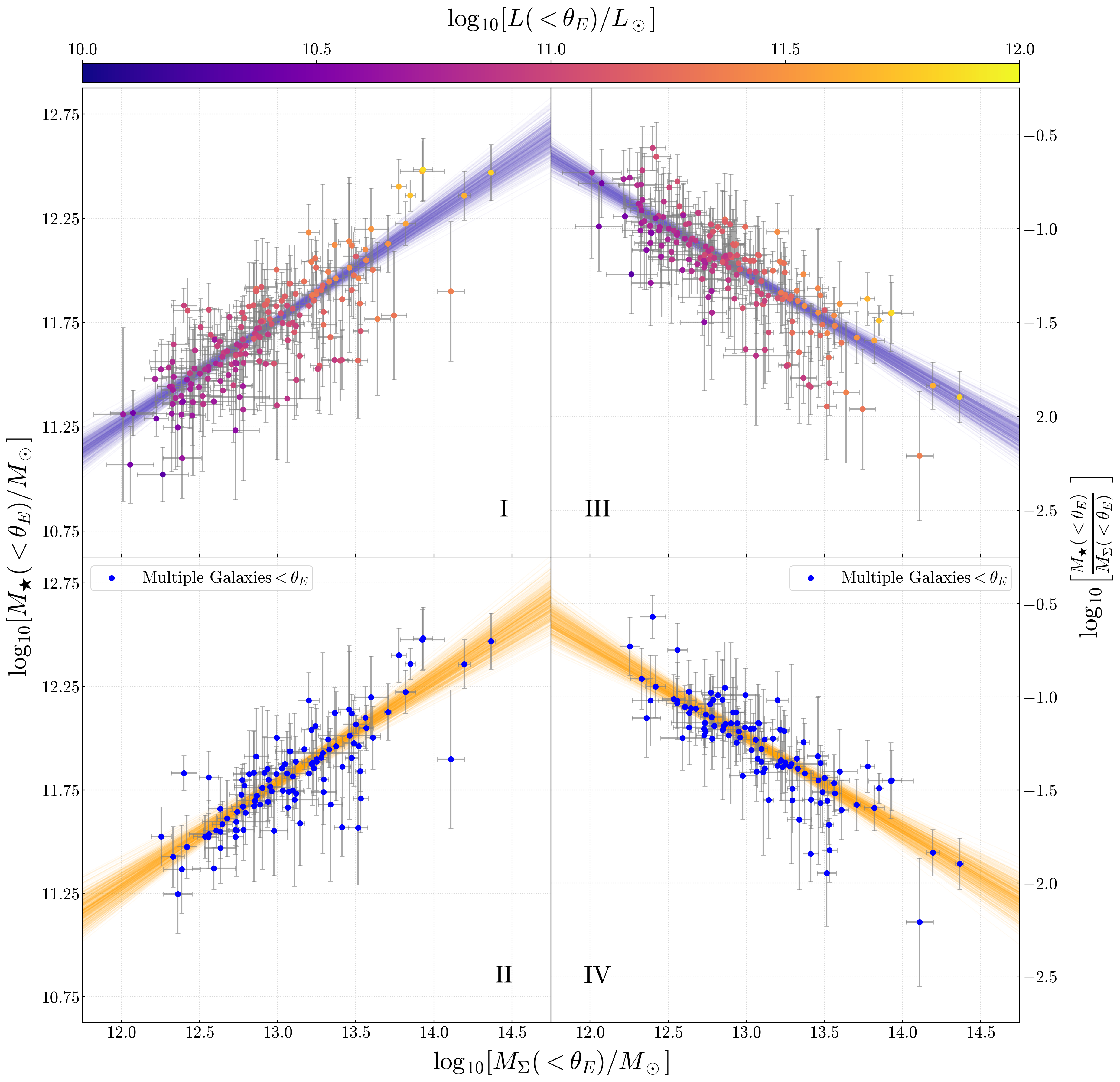}
    \caption{The total mass enclosed within the Einstein aperture for each system on the $x$-axis plotted against both the enclosed stellar mass and the enclosed stellar-to-total mass fraction on the $y$-axes. Points in Panels I and III are color coded according to the luminosity of the system in units of solar luminosities from stellar light enclosed within the Einstein aperture as measured within the rest-frame wavelength interval of 3000--7000\r{A}. Points in Panels II and IV only represent systems in which multiple cluster members fall within the Einstein aperture. Linear regression lines in all four Panels are drawn from the posterior distributions of their respective MCMC.}
    \label{four-for-four}
\end{figure*}

\section{Discussion}

\subsection{Scaling Relations}
In addition to the enclosed total mass and luminosity, we can also infer the observed enclosed stellar mass at the redshift of observation for each system. The stellar-mass posterior distribution from each SED constrained with \texttt{Prospector} describes the total amount of stellar mass formed over the lifetime of each system. However, the amount of stellar mass that we see at the redshift of observation is less than the total stellar mass formed over the lifetime of the system (e.g., such as supernovae or outflows; \citealt{2017ApJ...841..101L}). To correct for this, \texttt{Prospector} also returns a surviving mass fraction at the conclusion of each fit that describes the percentage of the total stellar mass that remains in the system at the redshift of observation. By multiplying this percentage by the best-fit stellar mass formed, we obtain the stellar mass at the redshift of observation for each system. While we note that enclosed stellar mass is more complex to infer and subject to more significant systematics when compared to the more direct measures of enclosed total mass from strong-lensing geometry and luminosity, it nevertheless offers another point of comparison with those measurables.

\subsubsection{Total Mass---Stellar Mass}
In Panel I of Figure \ref{four-for-four}, we see a clearly positive correlation between the enclosed total mass and the enclosed stellar mass for all systems. By performing a linear regression with \texttt{emcee} that accounts for errors in data on both axes, we find that the slope of this correlation is $0.500^{+0.029}_{-0.031}$ with an intrinsic scatter of $0.104^{+0.005}_{-0.003}$ dex from the posterior distribution of the MCMC fit. We colored the points according to their luminosity to highlight how the intrinsic differences between individual systems contribute to the overall intrinsic scatter. In Table 2 of \cite{2018AstL...44....8K}, they found $M_{500}$  \citep{2001A&A...367...27W} and the total stellar mass of the cluster within $R_{500}$ to be correlated with a logarithmic slope of $0.59\pm0.08$---in agreement with the slope we derive. 

\begin{figure}[!ht]
    \centering
    \includegraphics[width=8.5cm]{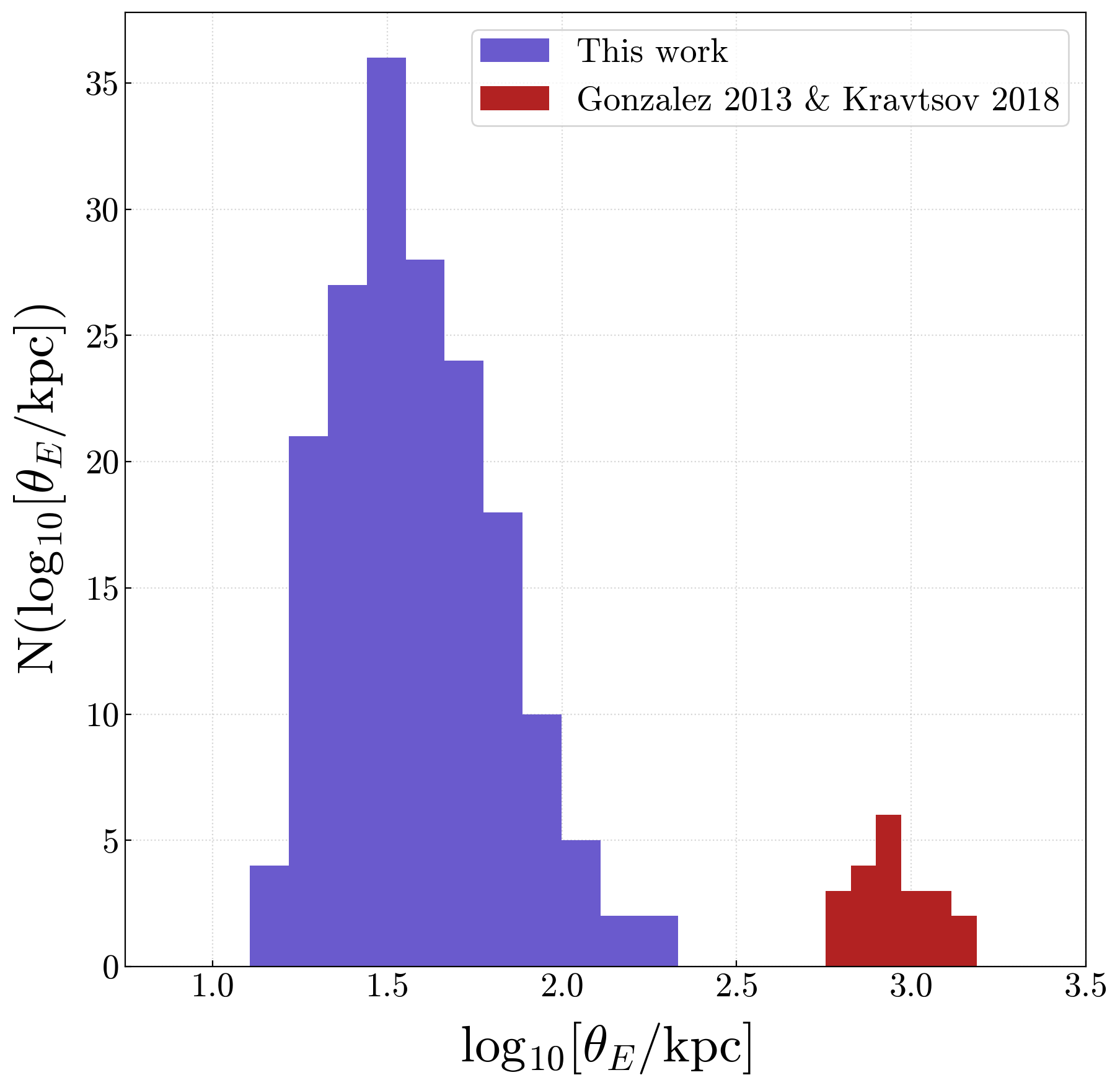}
    \caption{A log-space histogram of the fitted Einstein radii for this work scaled to physical units (kpc) shown in blue. Also shown is a log-space histogram of $r_{500}$ for 21 galaxy clusters analyzed in \cite{2018AstL...44....8K} and \cite{2013ApJ...778...14G}.}
    \label{einstein_andrey}
\end{figure}

Figure \ref{einstein_andrey} shows a distribution of the physical scales within which mass and light was measured for this work and that of \cite{2018AstL...44....8K} and \cite{2013ApJ...778...14G}. It is striking that, despite differentiating between the small-scale apertures considered in this work as well as large-scale apertures with a physical distance scale up to 30 times larger, we measure an identical relationship across clusters in that the more-massive clusters are less efficient in converting baryons into stars as indicated by a sub-unity stellar-to-total mass slope. We caution against overinterpretation of this result given that, in both cases, the total and stellar masses are cylindrical masses projected through a 3D halo. Therefore, even a very small aperture samples some mass and galaxy light at large radii in projection. Nevertheless, this result suggests that cluster galaxies at a range of cluster-centric radii, at least those probed by the combination of our results and \cite{2018AstL...44....8K} and \cite{2013ApJ...778...14G}, follow similar efficiency dependencies as a function of total halo mass. In addition, since cluster galaxies at larger radii are located within larger apertures, yet retain the same slope between stellar and total mass, cluster galaxies in the core and the outskirts of the cluster (i.e. at a range of cluster-centric radii) must be similarly affected by the mass of the cluster at large in terms of the total baryon conversion efficiency. Panel II of Figure \ref{four-for-four} further explores this regime, in which only those clusters where multiple cluster galaxies are contained within the Einstein aperture are plotted. We note that 103 of the clusters considered in this work met this criterion, and a linear regression for these systems gives a slope of $0.498^{+0.040}_{-0.039}$---identical to that of our entire sample.

Seeing as how the measured slopes agree across approximately two decades of cluster-centric radii suggests that whatever mechanism is responsible for suppressing baryon conversion efficiency is cluster wide and operating similarly regardless of the subpopulation of cluster member galaxies considered. What is unclear is whether this is predominantly a case of nature or nurture. Do the smaller halos that merge to form a massive cluster at later times have intrinsically lower baryon conversion efficiency, and that characteristic is inherited by the merged object (nature)? Or, is this suppression occurring at later times as a consequence of the halo growth through accretion and mergers (nurture)? That the stellar populations in cluster galaxies are much older than the assembled cluster-scale halo (e.g., \citealt{2022ApJ...934..177K}) suggests the former.

\subsubsection{Total Mass---Stellar-to-total-mass Fraction}

\cite{2010MNRAS.407..263A} measured 52 clusters and found that the stellar mass fraction depended on $M_{200}$ with a slope of $-0.55\pm0.08$. To compare directly with \cite{2010MNRAS.407..263A}, we plot the enclosed stellar-to-total mass fraction against the enclosed total mass in Panel III of Figure \ref{four-for-four} and see a clearly negative correlation (again with the same color-mapping as Panel I to emphasize intrinsic scatter). Also using \texttt{emcee}, we find that this correlation has a slope of $-0.495^{+0.032}_{-0.033}$ and an intrinsic scatter of $0.103^{+0.005}_{-0.002}$ dex. Categorizing in Panel IV identically to Panel II, we find that the multiple-galaxy systems are correlated with a slope of $-0.497^{+0.043}_{-0.047}$ and intrinsic scatter of $0.105^{+0.007}_{-0.004}$ dex. We again find that the slope for the multiple-galaxy systems is statistically identical to the slope for the entire sample shown in Panel III. The slopes we derive in Panels III and IV are statistically indistinguishable from those of \cite{2010MNRAS.407..263A}, who measured out to the physical distance scale of $R_{200}$---systematically larger than $R_{500}$ used in \cite{2018AstL...44....8K} and \cite{2013ApJ...778...14G} and much larger than this work.

While a later paper by \cite{2012A&A...548A..83A} found a shallower slope of $-0.38\pm0.07$ with a much smaller sample of only 12 galaxy clusters, the slope we derive with 177 clusters and that of \cite{2010MNRAS.407..263A} supports the prescription of a slope of $\sim-0.49$ from \cite{2009ApJ...700..989B} to reproduce observed X-ray scaling relations without any supplemental feedback from nonstellar sources. \cite{2018AstL...44....8K} also noted that the steepness of the slope they derived, in agreement with our own, indicates weaker feedback than cosmological simulations (e.g., \citealt{2005ApJ...625..588K, 2014MNRAS.440.2290M}) suggest. While this conclusion was not explicitly probed in this work, the large difference between the aperture sizes we use and those of \cite{2010MNRAS.407..263A} and \cite{2018AstL...44....8K} and the similarity of the derived slopes is suggestive of scale invariance with respect to feedback strength in addition to baryon conversion efficiency across a wide range of cluster-centric radii.

\subsection{Photometric Bias and Error}
\label{photo_err}
Deriving light in Section \ref{sec_photometry} via directly integrating over the Einstein aperture presents some challenges. Since we identified red-sequence cluster members by using observed color as a proxy for cluster membership, it is possible that we may have included intrinsically bluer galaxies at redshift $z>z_{_{BCG}}$ that appear to be the same color as the cluster members. Conversely, we may have excluded intrinsically bluer \textit{bona fide} cluster members. However, star-forming cluster members have less stellar mass and are intrinsically rarer in cluster cores (e.g., \citealt{1997ApJ...490..577D, 2004ApJ...617..867D}), and we do not expect that these missed cluster galaxies introduce a significant loss to the measured cluster light. In addition, bluer galaxies at redshift $z>z_{_{BCG}}$ appear relatively fainter, which are also unlikely to bias our methodology of measuring cluster light. Interlopers such as foreground stars that appear in front of cluster galaxies are forced to be masked, but any underlying cluster galaxy flux is typically minimal due to the limited angular size of such stars. 

In general, we expect that all of the unmasked flux from cluster cores at the angular scales considered in this work originates from genuine cluster members. Any modest background correction that could be made will also be effectively captured in the redshift dependence of the fits (e.g., Figure \ref{3d_plane_plot}). While we must also consider associated galaxy populations in the immediate large-scale structure when considering galaxy clusters projected onto the sky---as such structures may also be projected onto the cluster core---those galaxies will also contribute to the lensing effect and will have their mass and light captured by the methodology presented here.

As shown in Figure \ref{einstein_zitrin} and quantified in Section \ref{Einstein Radius}, our sample becomes incomplete at a measured Einstein radius of $3.09^{\prime\prime}$---falling to 50\% incomplete by $2.11^{\prime\prime}$. To some extent, this is a consequence of our chosen sample of cluster-scale lenses, which we would generally expect to have larger Einstein radii. However, we caution that there exists a large scatter between the total halo mass and Einstein radius \citep{2022ApJ...928...87F}. Systems whose arcs are separated from the central galaxy at small radii may also be located directly within the discernible BCG light. This would not only make them hard to visually notice, but it would also make them unsuitable for this analysis because the source light would be significantly blended together with the lensing cluster light. While we thus expect and demonstrate that this work undersamples the small-Einstein-radius regime, it is not apparent that this significantly alters our findings. 

In addition, our use of relatively shallow ground-based imaging limits the $5\sigma$ point-source magnitude depth of objects in the $g$-, $r$-, and $z$-band filters to $\approx$ 24.7, 23.9, and 23.0, respectively in the average case of two exposures per filter \citep{2019AJ....157..168D}. Systems with a lens redshift $z\gtrapprox1$ are barely visible even in $\mathit{z}$-band LS DR9 imaging data, and they are also simply a rarer type of strong lens in the Universe \citep{2019ApJ...878..122L}. Both the data depth and intrinsic redshift distribution of lenses shape the redshift distribution of the clusters in this work.

\subsection{Alternative Mass Measurement} \label{lenstool}
In constraining mass, we also considered a second method for measuring cluster-centric mass elucidated in \cite{2021ApJ...910..146R} using the parametric lens-modeling software \texttt{LENSTOOL} \citep{2007NJPh....9..447J} to measure the enclosed total mass---as opposed to the simple evaluation of Equations \eqref{mass_eq_1}, \eqref{mass_eq_2}, and \eqref{mass_eq_3}. For the first 35 clusters that were analyzed in this work, we generated single-halo lens models by taking the three coordinates in each lensed arc used to derive the Einstein radius in Section \ref{Einstein Radius} as a multiple-image family with a single pseudo-isothermal elliptical mass distribution \citep{1993ApJ...417..450K} with a free ellipticity locked to the center of the BCG. Mass estimates were then obtained from the resulting best-fit lens models by integrating over the surface-mass-density image, a 2D distribution in the lens plane of the mass fit with LENSTOOL. In this image, each pixel value represents the number of solar masses contained within that pixel. A circular aperture centered on the BCG with area equal to the area enclosed by the best-fit tangential critical curves was used to solve for the LENSTOOL enclosed mass in keeping with \cite{2021ApJ...910..146R}.

\begin{figure}[!ht]
    \centering
    \includegraphics[width=8.5cm]{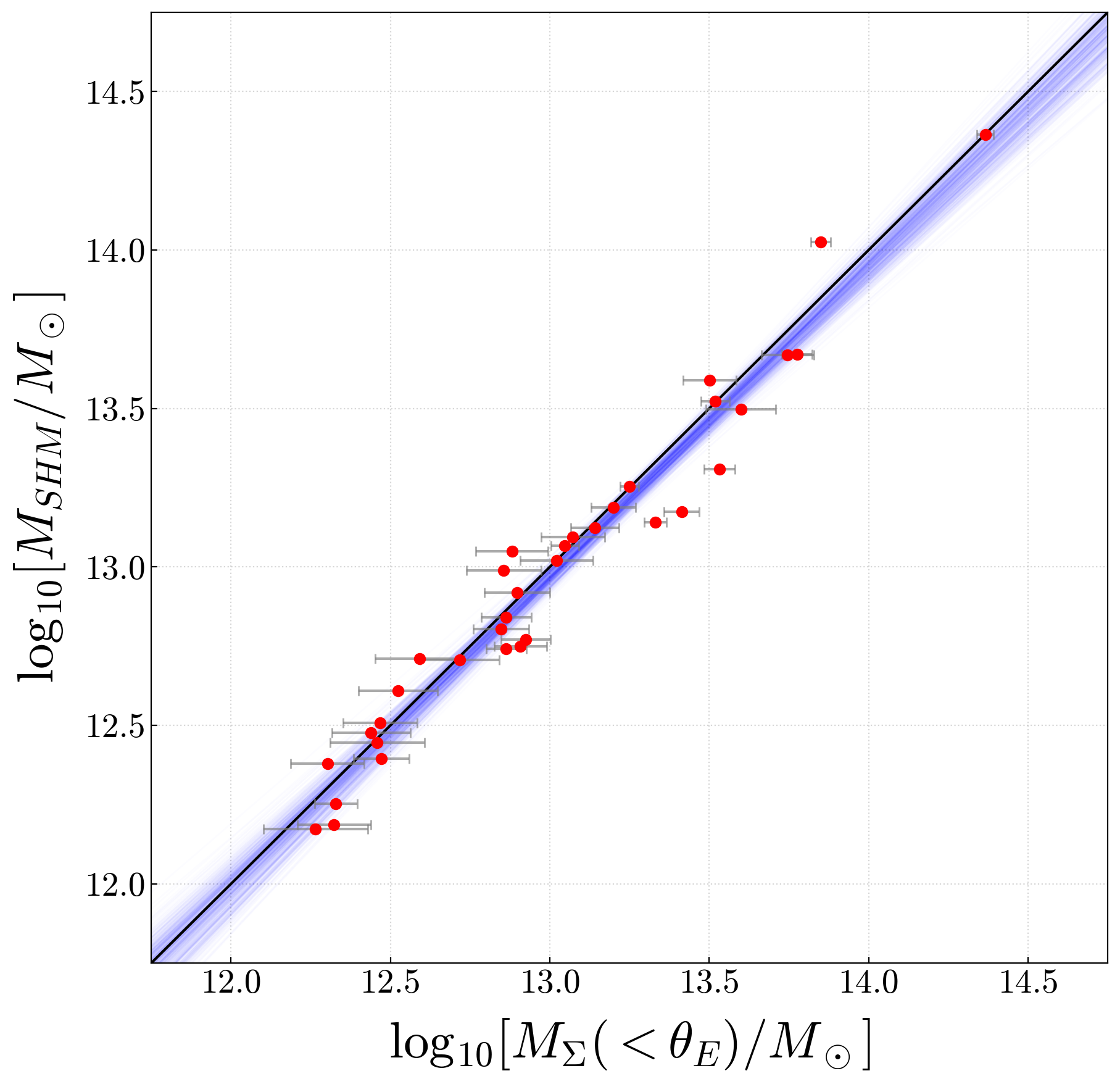}
    \caption{A comparison of the single-halo \texttt{LENSTOOL}-generated mass measurements ($M_{SHM}$) and the functionally generated mass measurements ($M_{\Sigma}$) for the first 35 clusters analyzed in this work. The black line represents an ideal one-to-one correspondence, and the blue lines represent the uncertainty on the measured correspondence.}
    \label{lenstool_comparison}
\end{figure}

As expected from \cite{2021ApJ...920...98R}, these differing methodologies show similar results. The two mass measurements for the 35 clusters studied here are correlated with a slope of $0.960\pm0.036$ and intercept of $0.492\pm0.476$ dex as shown in Figure \ref{lenstool_comparison}. In other words, we find no reason to suggest a preference between one method over the other, and we abandoned the \texttt{LENSTOOL} approach in favor of the much faster functional approach throughout the rest of the work. The agreement between the free-ellipticity-model masses and the circularly symmetric functional masses also demonstrates the feasibility of using a functional mass model for the geometrically simple strong lenses considered in this work.

Lens modeling has the potential to be the most accurate mass-constraining methodology in galaxy clusters. However, this requires high-spatial-resolution imaging data and ample time. The process of iteratively refining just one lens model may take upward of an hour for even a single relatively simple system---which is untenable when the number of candidate systems reaches into the thousands. Moreover, the accuracy of measuring mass with lens modeling is sensitive to the choice of defining multiple-image families along the lensed source arcs in a way that simple circle fitting is not. The essential positional constraints that define these image families are often unresolved---even with the highest-quality ground-based imaging data. Conversely, the positional constraints that inform the Einstein radius ultimately used by Equations \eqref{mass_eq_1}, \eqref{mass_eq_2}, and \eqref{mass_eq_3} need not be extremely accurate with respect to exactly constraining image family positions. As long as the positional constraints are reasonably placed within the arcs, something that \textit{is} resolved by ground-based imaging data, it is accurate to within the systematic uncertainty. 

For more information on lens-modeling software, related techniques, and applications strong lensing, we refer the reader to the following non-exhaustive list of papers for further review: \cite{2007arXiv0710.5636E}, \cite{2010PASJ...62.1017O}, \cite{2013NewAR..57....1L}, \cite{2017MNRAS.472.3177M}, \cite{2018PDU....22..189B}, \cite{2020A&C....3000360S}, \cite{2022ApJ...941..203S}, and \cite{2023ApJ...959..134N}.

\section{Summary and Conclusions}
This paper presents measurements of mass and light in 177 high-confidence cluster-scale strong lenses. By directly obtaining the BCG redshift and an estimate for the Einstein radius for each system, we compute the Einstein-radius-enclosed total mass functionally and independent of any light-traces-mass model. We use the enclosed $g$-, $r$-, and $z$-band LS DR9 photometry with \texttt{Prospector} to estimate the enclosed luminosity and stellar mass, and we obtain a planar relationship between the BCG redshift, enclosed total mass, and enclosed luminosity of the form $Z = 0.148^{+0.051}_{-0.052}X + 0.529^{+0.023}_{-0.022}Y + 4.121^{+0.272}_{-0.298}$ with an intrinsic scatter on $Z$ of $0.110^{+0.009}_{-0.008}$ dex where $X = z_{_{BCG}}$, $Y = \mathrm{log}_{10}[M_{\Sigma} (<\theta_E) / M_{\odot}]$, and $Z = \mathrm{log}_{10}[L(<\theta_{E})/L_\odot]$. We find that the enclosed total mass and stellar mass for the entire sample of strong lenses considered in this work are correlated with a logarithmic slope, normalization, and intrinsic stellar-mass scatter of $0.500^{+0.029}_{-0.031}$, $5.254^{+0.395}_{-0.391}$, $0.104^{+0.005}_{-0.003}$ dex, respectively. We also find that the enclosed total mass and stellar-to-total mass fraction are correlated with a logarithmic slope, normalization, and intrinsic stellar-to-total-mass-fraction scatter of $-0.495^{+0.032}_{-0.033}$, $5.24^{+0.409}_{-0.424}$, $0.103^{+0.005}_{-0.002}$ dex, respectively. We recover scaling relations identical to that of works that measure within cluster-scale apertures---despite the fact that we measure inside apertures with radii some 30--100 times smaller. This is suggestive of invariance in baryon conversion efficiency and feedback strength as a function of cluster-centric radii in galaxy clusters. Enormous volumes of imaging data from new space-based and ground-based surveys at deeper and sharper photometric limits are on the horizon. Probing these upcoming datasets with both visual and machine-learning-driven searches for new strong lenses at a variety of scales is inevitable. The correlations shown in this work should also have utility in ranking candidate strong lenses as well as offering another measure of cluster cores that can be compared to cosmological simulations.

\section*{Acknowledgments}
The authors would like to thank the anonymous reviewer for providing detailed feedback, which significantly improved the manuscript. S.D.M. would also like to thank Allison Noble for insightful discussions surrounding the  results and discussion Sections of this work.

This work is supported by The College undergraduate program and the College Innovation Fund at the University of Chicago and the Department of Astronomy and Astrophysics at the University of Chicago.


The Legacy Surveys consist of three individual and complementary projects: the Dark Energy Camera Legacy Survey (DECaLS; Proposal ID \#2014B-0404; PIs: David Schlegel and Arjun Dey), the Beijing-Arizona Sky Survey (BASS; NOAO Prop. ID \#2015A-0801; PIs: Zhou Xu and Xiaohui Fan), and the Mayall z-band Legacy Survey (MzLS; Prop. ID \#2016A-0453; PI: Arjun Dey). DECaLS, BASS and MzLS together include data obtained, respectively, at the Blanco telescope, Cerro Tololo Inter-American Observatory, NSF’s NOIRLab; the Bok telescope, Steward Observatory, University of Arizona; and the Mayall telescope, Kitt Peak National Observatory, NOIRLab. Pipeline processing and analyses of the data were supported by NOIRLab and the Lawrence Berkeley National Laboratory (LBNL). The Legacy Surveys project is honored to be permitted to conduct astronomical research on Iolkam Du’ag (Kitt Peak), a mountain with particular significance to the Tohono O’odham Nation.

NOIRLab is operated by the Association of Universities for Research in Astronomy (AURA) under a cooperative agreement with the National Science Foundation. LBNL is managed by the Regents of the University of California under contract to the U.S. Department of Energy.

This project used data obtained with the Dark Energy Camera (DECam), which was constructed by the Dark Energy Survey (DES) collaboration. Funding for the DES Projects has been provided by the U.S. Department of Energy, the U.S. National Science Foundation, the Ministry of Science and Education of Spain, the Science and Technology Facilities Council of the United Kingdom, the Higher Education Funding Council for England, the National Center for Supercomputing Applications at the University of Illinois at Urbana-Champaign, the Kavli Institute of Cosmological Physics at the University of Chicago, Center for Cosmology and Astro-Particle Physics at the Ohio State University, the Mitchell Institute for Fundamental Physics and Astronomy at Texas A\&M University, Financiadora de Estudos e Projetos, Fundacao Carlos Chagas Filho de Amparo, Financiadora de Estudos e Projetos, Fundacao Carlos Chagas Filho de Amparo a Pesquisa do Estado do Rio de Janeiro, Conselho Nacional de Desenvolvimento Cientifico e Tecnologico and the Ministerio da Ciencia, Tecnologia e Inovacao, the Deutsche Forschungsgemeinschaft and the Collaborating Institutions in the Dark Energy Survey. The Collaborating Institutions are Argonne National Laboratory, the University of California at Santa Cruz, the University of Cambridge, Centro de Investigaciones Energeticas, Medioambientales y Tecnologicas-Madrid, the University of Chicago, University College London, the DES-Brazil Consortium, the University of Edinburgh, the Eidgenossische Technische Hochschule (ETH) Zurich, Fermi National Accelerator Laboratory, the University of Illinois at Urbana-Champaign, the Institut de Ciencies de l’Espai (IEEC/CSIC), the Institut de Fisica d’Altes Energies, Lawrence Berkeley National Laboratory, the Ludwig Maximilians Universitat Munchen and the associated Excellence Cluster Universe, the University of Michigan, NSF’s NOIRLab, the University of Nottingham, the Ohio State University, the University of Pennsylvania, the University of Portsmouth, SLAC National Accelerator Laboratory, Stanford University, the University of Sussex, and Texas A\&M University.

BASS is a key project of the Telescope Access Program (TAP), which has been funded by the National Astronomical Observatories of China, the Chinese Academy of Sciences (the Strategic Priority Research Program “The Emergence of Cosmological Structures” Grant \# XDB09000000), and the Special Fund for Astronomy from the Ministry of Finance. The BASS is also supported by the External Cooperation Program of Chinese Academy of Sciences (Grant \# 114A11KYSB20160057), and Chinese National Natural Science Foundation (Grant \# 12120101003, \# 11433005).

The Legacy Survey team makes use of data products from the Near-Earth Object Wide-field Infrared Survey Explorer (NEOWISE), which is a project of the Jet Propulsion Laboratory/California Institute of Technology. NEOWISE is funded by the National Aeronautics and Space Administration.

The Legacy Surveys imaging of the DESI footprint is supported by the Director, Office of Science, Office of High Energy Physics of the U.S. Department of Energy under Contract No. DE-AC02-05CH1123, by the National Energy Research Scientific Computing Center, a DOE Office of Science User Facility under the same contract; and by the U.S. National Science Foundation, Division of Astronomical Sciences under Contract No. AST-0950945 to NOAO.

The Photometric Redshifts for the Legacy Surveys (PRLS) catalog used in this paper was produced thanks to funding from the U.S. Department of Energy Office of Science, Office of High Energy Physics via grant DE-SC0007914.


Funding for the Sloan Digital Sky Survey V has been provided by the Alfred P. Sloan Foundation, the Heising-Simons Foundation, the National Science Foundation, and the Participating Institutions. SDSS acknowledges support and resources from the Center for High-Performance Computing at the University of Utah. The SDSS web site is \url{www.sdss.org}.

SDSS is managed by the Astrophysical Research Consortium for the Participating Institutions of the SDSS Collaboration, including the Carnegie Institution for Science, Chilean National Time Allocation Committee (CNTAC) ratified researchers, the Gotham Participation Group, Harvard University, Heidelberg University, The Johns Hopkins University, L’Ecole polytechnique f\'{e}d\'{e}rale de Lausanne (EPFL), Leibniz-Institut f\"{u}r Astrophysik Potsdam (AIP), Max-Planck-Institut f\"{u}r Astronomie (MPIA Heidelberg), Max-Planck-Institut f\"{u}r Extraterrestrische Physik (MPE), Nanjing University, National Astronomical Observatories of China (NAOC), New Mexico State University, The Ohio State University, Pennsylvania State University, Smithsonian Astrophysical Observatory, Space Telescope Science Institute (STScI), the Stellar Astrophysics Participation Group, Universidad Nacional Aut\'{o}noma de M\'{e}xico, University of Arizona, University of Colorado Boulder, University of Illinois at Urbana-Champaign, University of Toronto, University of Utah, University of Virginia, Yale University, and Yunnan University.

\facilities{Sloan (BOSS spectrograph), Blanco (DECam wide-field camera)}

\software{
\texttt{astropy} \citep{astropy:2013, astropy:2018, astropy:2022}, 
\texttt{astro-prospector} \citep{2021ApJS..254...22J}, 
\texttt{emcee} \citep{2013PASP..125..306F},
\texttt{FSPS} \citep{2009ApJ...699..486C, 2010ApJ...712..833C},
\texttt{Jupyter Notebook} \citep{Kluyver2016jupyter}, 
\texttt{LENSTOOL} \citep{2007NJPh....9..447J}, 
\texttt{matplotlib} \citep{Hunter:2007}, 
\texttt{numpy} \citep{harris2020array}, 
\texttt{pandas} \citep{mckinney-proc-scipy-2010}, 
\texttt{python-FSPS} \citep{ben_johnson_2023_10026684}, 
\texttt{SAOImageDS9} \citep{2003ASPC..295..489J}, 
\texttt{scipy} \citep{2020SciPy-NMeth}.
}

\bibliography{references}{}
\bibliographystyle{aasjournal}

\startlongtable
\begin{deluxetable*}{lrrlrrrr}
\tabletypesize{\scriptsize}
\tablecolumns{8}
\tablecaption{Measured Quantities for All 177 Systems \label{tab:sample}}
\tablehead{\colhead{ID} & \colhead{RA [$^{\circ}$]} & \colhead{Dec [$^{\circ}$]} & \colhead{$z_{_{BCG}}$} & \colhead{$\theta_E$ [$^{\prime\prime}$]} & \colhead{$M_{\odot, \Sigma}$ [dex]} & \colhead{$M_{\odot, \star}$ [dex]} & \colhead{$L_{\odot}$ [dex]}}
\startdata 
J0008$+$1822 $^{\mathrm{m}}$ & 2.029829 & 18.372282 & 0.5566 $\pm$ 0.0483 $^{\mathrm{p}}$ & 3.2840 $\pm$ 0.3121 & 12.3613 $\pm$ 0.0912 & 11.2470 $\pm$ 0.1891 & 10.5487 $\pm$ 0.0947 \\
J0008$-$0624 $^{\mathrm{o}}$ & 2.169667 & -6.405227 & 0.9801 $\pm$ 0.0584 $^{\mathrm{p}}$ & 3.7392 $\pm$ 0.3104 & 12.7804 $\pm$ 0.0922 & 11.3325 $\pm$ 0.3413 & 10.7884 $\pm$ 0.0757 \\
J0008$+$2150 $^{\mathrm{m}}$ & 2.181223 & 21.836401 & 0.5974 $\pm$ 0.0190 $^{\mathrm{p}}$ & 7.3700 $\pm$ 0.5581 & 13.0964 $\pm$ 0.0703 & 11.7489 $\pm$ 0.1468 & 11.0052 $\pm$ 0.0389 \\
J0019$+$0438 $^{\mathrm{m}}$ & 4.954375 & 4.649355 & 0.8685 $\pm$ 0.0339 $^{\mathrm{p}}$ & 5.8629 $\pm$ 1.0055 & 13.1077 $\pm$ 0.1544 & 11.7035 $\pm$ 0.4180 & 11.1106 $\pm$ 0.0938 \\
J0021$+$0333 $^{\mathrm{m}}$ & 5.402546 & 3.556039 & 0.7164 $\pm$ 0.0222 $^{\mathrm{p}}$ & 4.5574 $\pm$ 0.3604 & 12.7671 $\pm$ 0.0760 & 11.7287 $\pm$ 0.1243 & 11.0022 $\pm$ 0.0429 \\
J0022$+$1431 $^{\mathrm{o}}_*$ & 5.670506 & 14.519566 & 0.3805 $\pm$ 0.0001 $^{\mathrm{s}}$ & 3.8494 $\pm$ 0.2942 & 12.3287 $\pm$ 0.0669 & 11.3592 $\pm$ 0.1336 & 10.7212 $\pm$ 0.0077 \\
J0023$-$0252 $^{\mathrm{o}}_*$ & 5.835225 & -2.874259 & 0.7973 $\pm$ 0.0698 $^{\mathrm{p}}$ & 2.0601 $\pm$ 0.3552 & 12.2650 $\pm$ 0.1635 & 11.0213 $\pm$ 0.1294 & 10.3088 $\pm$ 0.0998 \\
J0027$-$0413 $^{\mathrm{m}}$ & 6.750411 & -4.223259 & 0.4950 $\pm$ 0.0001 $^{\mathrm{s}}$ & 5.6319 $\pm$ 0.3836 & 12.7744 $\pm$ 0.0603 & 11.7977 $\pm$ 0.0768 & 11.0663 $\pm$ 0.0076 \\
J0030$-$0101 $^{\mathrm{o}}$ & 7.558161 & -1.029405 & 0.7234 $\pm$ 0.0954 $^{\mathrm{p}}$ & 3.7474 $\pm$ 0.5409 & 12.7308 $\pm$ 0.1511 & 11.2329 $\pm$ 0.3320 & 10.5886 $\pm$ 0.1750 \\
J0032$+$0740 $^{\mathrm{o}}$ & 8.013587 & 7.667835 & 0.6437 $\pm$ 0.0003 $^{\mathrm{s}}$ & 3.1170 $\pm$ 0.3689 & 12.5041 $\pm$ 0.1046 & 11.5044 $\pm$ 0.2773 & 10.7852 $\pm$ 0.0271 \\
J0034$+$0225 $^{\mathrm{m}}_*$ & 8.617335 & 2.422958 & 0.3865 $\pm$ 0.0077 $^{\mathrm{p}}$ & 13.3401 $\pm$ 0.8185 & 13.4137 $\pm$ 0.0548 & 11.8617 $\pm$ 0.1459 & 11.2455 $\pm$ 0.0213 \\
J0036$-$0506 $^{\mathrm{o}}$ & 9.081628 & -5.108834 & 0.8121 $\pm$ 0.0223 $^{\mathrm{p}}$ & 3.9209 $\pm$ 0.3094 & 12.7063 $\pm$ 0.0787 & 11.5498 $\pm$ 0.2415 & 10.9275 $\pm$ 0.0361 \\
J0038$+$0719 $^{\mathrm{m}}$ & 9.730731 & 7.323072 & 0.2643 $\pm$ 0.0178 $^{\mathrm{p}}$ & 7.3875 $\pm$ 0.2963 & 12.7385 $\pm$ 0.0452 & 11.5541 $\pm$ 0.0753 & 10.7391 $\pm$ 0.0714 \\
J0046$-$0156 $^{\mathrm{m}}$ & 11.508417 & -1.941216 & 0.5501 $\pm$ 0.0002 $^{\mathrm{s}}$ & 5.3411 $\pm$ 0.4457 & 12.7774 $\pm$ 0.0733 & 11.6689 $\pm$ 0.1161 & 10.9302 $\pm$ 0.0106 \\
J0047$+$0508 $^{\mathrm{m}}$ & 11.961840 & 5.138756 & 0.4291 $\pm$ 0.0001 $^{\mathrm{s}}$ & 8.9939 $\pm$ 0.3326 & 13.1163 $\pm$ 0.0339 & 11.8872 $\pm$ 0.1224 & 11.1796 $\pm$ 0.0120 \\
J0051$-$0923 $^{\mathrm{o}}$ & 12.900786 & -9.390354 & 0.7134 $\pm$ 0.0243 $^{\mathrm{p}}$ & 3.6429 $\pm$ 0.3584 & 12.5724 $\pm$ 0.0898 & 11.6910 $\pm$ 0.1329 & 11.0588 $\pm$ 0.0424 \\
J0056$-$1402 $^{\mathrm{m}}$ & 14.247809 & -14.042115 & 0.4251 $\pm$ 0.0208 $^{\mathrm{p}}$ & 6.1293 $\pm$ 0.4406 & 12.7821 $\pm$ 0.0676 & 11.5572 $\pm$ 0.1479 & 10.8989 $\pm$ 0.0590 \\
J0057$+$2138 $^{\mathrm{o}}$ & 14.312163 & 21.642771 & 0.8471 $\pm$ 0.0402 $^{\mathrm{p}}$ & 4.0709 $\pm$ 0.2746 & 12.7619 $\pm$ 0.0699 & 11.5992 $\pm$ 0.1372 & 10.9691 $\pm$ 0.0566 \\
J0057$+$0230 $^{\mathrm{o}}$ & 14.421865 & 2.504200 & 0.9262 $\pm$ 0.0005 $^{\mathrm{s}}$ & 3.8683 $\pm$ 0.3260 & 12.7703 $\pm$ 0.0823 & 11.6276 $\pm$ 0.1559 & 11.0628 $\pm$ 0.0209 \\
J0058$-$0721 $^{\mathrm{m}}_*$ & 14.703994 & -7.365765 & 0.6379 $\pm$ 0.0001 $^{\mathrm{s}}$ & 15.5864 $\pm$ 0.7642 & 13.7757 $\pm$ 0.0475 & 12.4018 $\pm$ 0.1310 & 11.6667 $\pm$ 0.0082 \\
J0100$+$1818 $^{\mathrm{o}}_*$ & 15.204935 & 18.307730 & 0.5819 $\pm$ 0.0002 $^{\mathrm{s}}$ & 6.8790 $\pm$ 0.5811 & 13.1414 $\pm$ 0.0756 & 11.7847 $\pm$ 0.1288 & 11.0149 $\pm$ 0.0087 \\
J0101$+$2055 $^{\mathrm{o}}$ & 15.437768 & 20.928614 & 0.8708 $\pm$ 0.0003 $^{\mathrm{s}}$ & 4.6041 $\pm$ 0.3144 & 12.8860 $\pm$ 0.0704 & 11.6122 $\pm$ 0.1915 & 10.8987 $\pm$ 0.0306 \\
J0101$-$2126 $^{\mathrm{o}}_*$ & 15.498254 & -21.448722 & 0.4001 $\pm$ 0.0375 $^{\mathrm{p}}$ & 4.4297 $\pm$ 0.3905 & 12.4722 $\pm$ 0.0872 & 11.6162 $\pm$ 0.1226 & 10.8636 $\pm$ 0.1010 \\
J0102$-$1400 $^{\mathrm{o}}$ & 15.676298 & -14.015030 & 0.6737 $\pm$ 0.0321 $^{\mathrm{p}}$ & 3.6148 $\pm$ 0.2946 & 12.6504 $\pm$ 0.0778 & 11.4241 $\pm$ 0.0921 & 10.8318 $\pm$ 0.0537 \\
J0102$-$2356 $^{\mathrm{o}}$ & 15.678569 & -23.943939 & 0.7084 $\pm$ 0.0286 $^{\mathrm{p}}$ & 3.1963 $\pm$ 0.2857 & 12.4547 $\pm$ 0.0824 & 11.3045 $\pm$ 0.1416 & 10.6862 $\pm$ 0.0549 \\
J0103$+$2706 $^{\mathrm{m}}$ & 15.937230 & 27.102542 & 0.3825 $\pm$ 0.0001 $^{\mathrm{s}}$ & 4.3371 $\pm$ 0.6150 & 12.5583 $\pm$ 0.1225 & 11.5363 $\pm$ 0.1660 & 10.7703 $\pm$ 0.0334 \\
J0104$-$0340 $^{\mathrm{o}}$ & 16.214582 & -3.671084 & 0.9163 $\pm$ 0.0350 $^{\mathrm{p}}$ & 6.9245 $\pm$ 0.2938 & 13.2689 $\pm$ 0.0612 & 11.5434 $\pm$ 0.2284 & 11.1627 $\pm$ 0.0486 \\
J0104$-$0757 $^{\mathrm{o}}_*$ & 16.220190 & -7.952046 & 0.8988 $\pm$ 0.0401 $^{\mathrm{p}}$ & 2.9478 $\pm$ 0.3867 & 12.5238 $\pm$ 0.1246 & 11.3588 $\pm$ 0.1544 & 10.8748 $\pm$ 0.0585 \\
J0105$-$0501 $^{\mathrm{o}}$ & 16.462922 & -5.032347 & 0.7015 $\pm$ 0.0002 $^{\mathrm{s}}$ & 3.7881 $\pm$ 0.2956 & 12.5959 $\pm$ 0.0696 & 11.6725 $\pm$ 0.1452 & 11.0599 $\pm$ 0.0089 \\
J0109$-$0455 $^{\mathrm{m}}$ & 17.294557 & -4.919535 & 0.7623 $\pm$ 0.0352 $^{\mathrm{p}}$ & 5.1884 $\pm$ 0.3166 & 12.9158 $\pm$ 0.0681 & 11.8335 $\pm$ 0.1731 & 11.1693 $\pm$ 0.0516 \\
J0109$-$3335 $^{\mathrm{m}}$ & 17.469499 & -33.592604 & 0.4730 $\pm$ 0.0114 $^{\mathrm{p}}$ & 9.1133 $\pm$ 0.2983 & 13.1705 $\pm$ 0.0333 & 11.9468 $\pm$ 0.0989 & 11.2857 $\pm$ 0.0253 \\
J0111$+$1346 $^{\mathrm{m}}$ & 17.779877 & 13.778901 & 0.7203 $\pm$ 0.0002 $^{\mathrm{s}}$ & 5.1998 $\pm$ 1.0656 & 12.9016 $\pm$ 0.1788 & 11.7595 $\pm$ 0.3941 & 11.1605 $\pm$ 0.0844 \\
J0111$+$0855 $^{\mathrm{m}}$ & 17.881205 & 8.928360 & 0.4855 $\pm$ 0.0001 $^{\mathrm{s}}$ & 11.2512 $\pm$ 0.4636 & 13.3656 $\pm$ 0.0389 & 12.1219 $\pm$ 0.1217 & 11.4547 $\pm$ 0.0129 \\
J0112$-$2053 $^{\mathrm{m}}_*$ & 18.014632 & -20.891104 & 0.6441 $\pm$ 0.0355 $^{\mathrm{p}}$ & 5.4066 $\pm$ 0.4312 & 12.8629 $\pm$ 0.0785 & 11.9117 $\pm$ 0.2331 & 11.2771 $\pm$ 0.0624 \\
J0117$-$0527 $^{\mathrm{m}}$ & 19.494867 & -5.454945 & 0.5797 $\pm$ 0.0002 $^{\mathrm{s}}$ & 2.8806 $\pm$ 0.3928 & 12.3875 $\pm$ 0.1171 & 11.3673 $\pm$ 0.2130 & 10.6406 $\pm$ 0.0479 \\
J0118$-$0526 $^{\mathrm{m}}$ & 19.663276 & -5.443448 & 0.5803 $\pm$ 0.0002 $^{\mathrm{s}}$ & 12.8524 $\pm$ 0.3093 & 13.5624 $\pm$ 0.0279 & 12.0997 $\pm$ 0.1497 & 11.4396 $\pm$ 0.0099 \\
J0122$-$0831 $^{\mathrm{o}}$ & 20.604365 & -8.520537 & 0.4906 $\pm$ 0.0495 $^{\mathrm{p}}$ & 2.1354 $\pm$ 0.3496 & 12.0570 $\pm$ 0.1503 & 11.0683 $\pm$ 0.1841 & 10.4664 $\pm$ 0.1147 \\
J0124$-$2401 $^{\mathrm{o}}$ & 21.086027 & -24.029980 & 0.6187 $\pm$ 0.0092 $^{\mathrm{p}}$ & 3.0023 $\pm$ 0.3254 & 12.3345 $\pm$ 0.0956 & 11.6448 $\pm$ 0.2119 & 10.9321 $\pm$ 0.0237 \\
J0127$+$2713 $^{\mathrm{m}}$ & 21.939205 & 27.229009 & 0.7989 $\pm$ 0.0234 $^{\mathrm{p}}$ & 7.1497 $\pm$ 0.4211 & 13.2163 $\pm$ 0.0610 & 12.0409 $\pm$ 0.1081 & 11.3445 $\pm$ 0.0335 \\
J0130$-$0305 $^{\mathrm{m}}$ & 22.517741 & -3.095758 & 0.6754 $\pm$ 0.1767 $^{\mathrm{p}}$ & 17.8150 $\pm$ 0.3962 & 13.9260 $\pm$ 0.1418 & 12.4753 $\pm$ 0.1461 & 11.6512 $\pm$ 0.3342 \\
J0130$+$3216 $^{\mathrm{o}}$ & 22.730273 & 32.275463 & 0.5093 $\pm$ 0.0001 $^{\mathrm{s}}$ & 9.5570 $\pm$ 0.9385 & 13.3655 $\pm$ 0.0844 & 11.5700 $\pm$ 0.1060 & 10.9152 $\pm$ 0.0106 \\
J0133$-$1650 $^{\mathrm{m}}$ & 23.423964 & -16.839019 & 0.6490 $\pm$ 0.0191 $^{\mathrm{p}}$ & 6.1292 $\pm$ 0.5836 & 12.9769 $\pm$ 0.0864 & 11.5527 $\pm$ 0.2162 & 11.0845 $\pm$ 0.0401 \\
J0134$+$0433 $^{\mathrm{o}}$ & 23.676569 & 4.563916 & 0.5509 $\pm$ 0.0001 $^{\mathrm{s}}$ & 5.3402 $\pm$ 0.3406 & 12.8939 $\pm$ 0.0587 & 11.7543 $\pm$ 0.0771 & 11.0184 $\pm$ 0.0079 \\
J0135$-$2033 $^{\mathrm{m}}_*$ & 23.928317 & -20.559902 & 0.6111 $\pm$ 0.0169 $^{\mathrm{p}}$ & 4.7636 $\pm$ 0.4582 & 12.8466 $\pm$ 0.0869 & 11.8326 $\pm$ 0.1995 & 11.1405 $\pm$ 0.0342 \\
J0136$-$2200 $^{\mathrm{o}}$ & 24.211948 & -22.007600 & 0.6741 $\pm$ 0.0247 $^{\mathrm{p}}$ & 2.6516 $\pm$ 0.2856 & 12.3852 $\pm$ 0.0992 & 11.3087 $\pm$ 0.2153 & 10.6889 $\pm$ 0.0639 \\
J0137$+$2109 $^{\mathrm{o}}$ & 24.385656 & 21.156756 & 0.8805 $\pm$ 0.0610 $^{\mathrm{p}}$ & 4.4855 $\pm$ 0.6013 & 12.9960 $\pm$ 0.1293 & 11.3528 $\pm$ 0.2593 & 10.9269 $\pm$ 0.0807 \\
J0138$-$2155 $^{\mathrm{m}}_*$ & 24.515695 & -21.925493 & 0.2468 $\pm$ 0.0451 $^{\mathrm{p}}$ & 18.3359 $\pm$ 0.5885 & 13.5015 $\pm$ 0.0833 & 12.0665 $\pm$ 0.0438 & 11.1369 $\pm$ 0.2022 \\
J0138$-$2844 $^{\mathrm{m}}$ & 24.597416 & -28.735794 & 0.4005 $\pm$ 0.0141 $^{\mathrm{p}}$ & 8.7874 $\pm$ 0.3093 & 13.0654 $\pm$ 0.0364 & 11.6649 $\pm$ 0.1556 & 10.9957 $\pm$ 0.0374 \\
J0139$+$2207 $^{\mathrm{o}}$ & 24.778355 & 22.123106 & 0.4770 $\pm$ 0.0001 $^{\mathrm{s}}$ & 5.8704 $\pm$ 0.5433 & 12.7958 $\pm$ 0.0805 & 11.5995 $\pm$ 0.1068 & 10.8991 $\pm$ 0.0102 \\
J0140$-$2006 $^{\mathrm{m}}$ & 25.052071 & -20.105520 & 0.3617 $\pm$ 0.0070 $^{\mathrm{p}}$ & 17.2884 $\pm$ 0.9038 & 13.6096 $\pm$ 0.0469 & 12.0021 $\pm$ 0.1522 & 11.3085 $\pm$ 0.0205 \\
J0143$+$1427 $^{\mathrm{o}}$ & 25.953318 & 14.461970 & 0.3305 $\pm$ 0.0001 $^{\mathrm{s}}$ & 6.9242 $\pm$ 0.4053 & 12.7779 $\pm$ 0.0515 & 11.4450 $\pm$ 0.1645 & 10.7016 $\pm$ 0.0051 \\
J0144$-$2213 $^{\mathrm{m}}$ & 26.168407 & -22.229454 & 0.2773 $\pm$ 0.0068 $^{\mathrm{p}}$ & 12.5641 $\pm$ 0.2836 & 13.2196 $\pm$ 0.0231 & 11.8759 $\pm$ 0.0506 & 11.0744 $\pm$ 0.0249 \\
J0144$+$1008 $^{\mathrm{o}}$ & 26.172042 & 10.141264 & 0.4649 $\pm$ 0.0001 $^{\mathrm{s}}$ & 4.2238 $\pm$ 0.3345 & 12.4975 $\pm$ 0.0703 & 11.4401 $\pm$ 0.1328 & 10.7675 $\pm$ 0.0127 \\
J0145$-$3541 $^{\mathrm{o}}$ & 26.444968 & -35.691008 & 0.4986 $\pm$ 0.0182 $^{\mathrm{p}}$ & 3.8242 $\pm$ 0.2939 & 12.4415 $\pm$ 0.0703 & 11.4313 $\pm$ 0.0862 & 10.7376 $\pm$ 0.0401 \\
J0145$+$0402 $^{\mathrm{o}}$ & 26.484783 & 4.041294 & 0.7800 $\pm$ 0.0410 $^{\mathrm{p}}$ & 2.3157 $\pm$ 0.2983 & 12.3479 $\pm$ 0.1176 & 11.3878 $\pm$ 0.3417 & 10.7814 $\pm$ 0.0674 \\
J0146$-$0929 $^{\mathrm{m}}_*$ & 26.733372 & -9.497905 & 0.4469 $\pm$ 0.0001 $^{\mathrm{s}}$ & 12.2287 $\pm$ 0.5991 & 13.5186 $\pm$ 0.0451 & 11.9612 $\pm$ 0.1286 & 11.4089 $\pm$ 0.0042 \\
J0149$-$2834 $^{\mathrm{o}}$ & 27.483033 & -28.575331 & 0.5563 $\pm$ 0.0076 $^{\mathrm{p}}$ & 5.9089 $\pm$ 0.2878 & 12.8687 $\pm$ 0.0461 & 11.6850 $\pm$ 0.0784 & 10.9697 $\pm$ 0.0167 \\
J0150$+$2725 $^{\mathrm{m}}$ & 27.503618 & 27.426752 & 0.3062 $\pm$ 0.0001 $^{\mathrm{s}}$ & 5.5610 $\pm$ 0.4277 & 12.5561 $\pm$ 0.0674 & 11.5224 $\pm$ 0.1160 & 10.7598 $\pm$ 0.0086 \\
J0151$-$0407 $^{\mathrm{m}}$ & 27.908847 & -4.120915 & 0.6700 $\pm$ 0.0002 $^{\mathrm{s}}$ & 6.8046 $\pm$ 0.3487 & 13.0807 $\pm$ 0.0511 & 11.9371 $\pm$ 0.1355 & 11.1499 $\pm$ 0.0118 \\
J0153$-$3235 $^{\mathrm{o}}$ & 28.474812 & -32.599633 & 0.6541 $\pm$ 0.0204 $^{\mathrm{p}}$ & 4.2235 $\pm$ 0.2766 & 12.6551 $\pm$ 0.0646 & 11.6037 $\pm$ 0.0940 & 10.9579 $\pm$ 0.0355 \\
J0159$-$3413 $^{\mathrm{m}}$ & 29.766637 & -34.217907 & 0.4143 $\pm$ 0.0252 $^{\mathrm{p}}$ & 10.4741 $\pm$ 0.3272 & 13.2318 $\pm$ 0.0393 & 11.8553 $\pm$ 0.1337 & 11.2137 $\pm$ 0.0628 \\
J0202$-$1109 $^{\mathrm{m}}_*$ & 30.543766 & -11.153215 & 0.4444 $\pm$ 0.0237 $^{\mathrm{p}}$ & 5.6661 $\pm$ 0.7555 & 12.8553 $\pm$ 0.1174 & 11.6974 $\pm$ 0.1213 & 10.9311 $\pm$ 0.0577 \\
J0203$-$2017 $^{\mathrm{m}}$ & 30.793373 & -20.289939 & 0.4389 $\pm$ 0.0227 $^{\mathrm{p}}$ & 13.4329 $\pm$ 0.5003 & 13.4748 $\pm$ 0.0413 & 11.9055 $\pm$ 0.1245 & 11.2181 $\pm$ 0.0558 \\
J0204$-$2918 $^{\mathrm{m}}$ & 31.034313 & -29.302323 & 0.9894 $\pm$ 0.0775 $^{\mathrm{p}}$ & 4.0037 $\pm$ 0.2630 & 12.8484 $\pm$ 0.0904 & 11.6719 $\pm$ 0.3641 & 11.1179 $\pm$ 0.1043 \\
J0205$-$3539 $^{\mathrm{m}}$ & 31.358759 & -35.663177 & 0.3856 $\pm$ 0.0101 $^{\mathrm{p}}$ & 3.5102 $\pm$ 0.2502 & 12.2543 $\pm$ 0.0641 & 11.5256 $\pm$ 0.1424 & 10.7422 $\pm$ 0.0287 \\
J0209$-$3547 $^{\mathrm{m}}$ & 32.476474 & -35.799023 & 0.3105 $\pm$ 0.0097 $^{\mathrm{p}}$ & 5.8467 $\pm$ 0.5346 & 12.6066 $\pm$ 0.0805 & 11.5537 $\pm$ 0.0804 & 10.7572 $\pm$ 0.0337 \\
J0210$+$2600 $^{\mathrm{o}}$ & 32.594801 & 26.011166 & 0.3960 $\pm$ 0.0001 $^{\mathrm{s}}$ & 3.3126 $\pm$ 0.2662 & 12.2154 $\pm$ 0.0702 & 11.4794 $\pm$ 0.1365 & 10.7399 $\pm$ 0.0098 \\
J0210$+$3044 $^{\mathrm{o}}$ & 32.744179 & 30.739885 & 0.4687 $\pm$ 0.1181 $^{\mathrm{p}}$ & 5.5452 $\pm$ 0.9363 & 12.7579 $\pm$ 0.1926 & 11.3927 $\pm$ 0.1687 & 10.7470 $\pm$ 0.3266 \\
J0214$-$0206 $^{\mathrm{o}}_*$ & 33.533392 & -2.107933 & 0.6477 $\pm$ 0.0322 $^{\mathrm{p}}$ & 2.9660 $\pm$ 0.3965 & 12.4676 $\pm$ 0.1169 & 11.5292 $\pm$ 0.2349 & 10.8141 $\pm$ 0.0683 \\
J0224$+$0849 $^{\mathrm{o}}_*$ & 36.233942 & 8.829935 & 0.3271 $\pm$ 0.0225 $^{\mathrm{p}}$ & 7.6755 $\pm$ 0.4873 & 12.8633 $\pm$ 0.0628 & 11.9000 $\pm$ 0.1489 & 11.1706 $\pm$ 0.0712 \\
J0225$-$0737 $^{\mathrm{m}}$ & 36.442200 & -7.627366 & 0.5137 $\pm$ 0.0001 $^{\mathrm{s}}$ & 6.2834 $\pm$ 0.5378 & 12.8870 $\pm$ 0.0758 & 11.6802 $\pm$ 0.1605 & 10.9907 $\pm$ 0.0164 \\
J0227$+$2934 $^{\mathrm{m}}_*$ & 36.806665 & 29.579592 & 0.4856 $\pm$ 0.0351 $^{\mathrm{p}}$ & 3.9821 $\pm$ 0.6265 & 12.5917 $\pm$ 0.1397 & 11.3713 $\pm$ 0.1022 & 10.6831 $\pm$ 0.0767 \\
J0228$-$2923 $^{\mathrm{o}}$ & 37.067992 & -29.396016 & 0.2951 $\pm$ 0.0106 $^{\mathrm{p}}$ & 3.8560 $\pm$ 0.3343 & 12.2240 $\pm$ 0.0773 & 11.2894 $\pm$ 0.0826 & 10.4594 $\pm$ 0.0406 \\
J0230$-$2702 $^{\mathrm{m}}$ & 37.702673 & -27.041580 & 0.3874 $\pm$ 0.0330 $^{\mathrm{p}}$ & 13.3091 $\pm$ 0.5163 & 13.4125 $\pm$ 0.0501 & 11.5702 $\pm$ 0.1413 & 10.9837 $\pm$ 0.0912 \\
J0233$+$0559 $^{\mathrm{m}}$ & 38.348967 & 5.999309 & 0.2688 $\pm$ 0.0122 $^{\mathrm{p}}$ & 13.0829 $\pm$ 0.3091 & 13.2417 $\pm$ 0.0292 & 12.0576 $\pm$ 0.1080 & 11.2895 $\pm$ 0.0502 \\
J0233$+$0642 $^{\mathrm{o}}$ & 38.488527 & 6.711545 & 0.5972 $\pm$ 0.0179 $^{\mathrm{p}}$ & 3.0194 $\pm$ 0.3165 & 12.3238 $\pm$ 0.0943 & 11.4435 $\pm$ 0.1297 & 10.8144 $\pm$ 0.0350 \\
J0237$-$3017 $^{\mathrm{m}}$ & 39.348438 & -30.292085 & 0.5451 $\pm$ 0.0182 $^{\mathrm{p}}$ & 5.4340 $\pm$ 0.3750 & 12.7875 $\pm$ 0.0646 & 11.7716 $\pm$ 0.1077 & 11.1021 $\pm$ 0.0363 \\
J0239$-$2047 $^{\mathrm{o}}$ & 39.776954 & -20.788370 & 0.6268 $\pm$ 0.0098 $^{\mathrm{p}}$ & 3.6512 $\pm$ 0.2932 & 12.5082 $\pm$ 0.0725 & 11.7279 $\pm$ 0.1370 & 10.9814 $\pm$ 0.0227 \\
J0239$-$0134 $^{\mathrm{m}}$ & 39.971351 & -1.582259 & 0.3731 $\pm$ 0.0001 $^{\mathrm{s}}$ & 11.2410 $\pm$ 0.7448 & 13.2506 $\pm$ 0.0584 & 11.8986 $\pm$ 0.1494 & 11.2838 $\pm$ 0.0081 \\
J0248$-$0216 $^{\mathrm{m}}$ & 42.034802 & -2.276981 & 0.2338 $\pm$ 0.0000 $^{\mathrm{s}}$ & 13.3466 $\pm$ 0.3547 & 13.2003 $\pm$ 0.0238 & 11.8296 $\pm$ 0.0725 & 11.0756 $\pm$ 0.0044 \\
J0251$-$1220 $^{\mathrm{m}}$ & 42.897099 & -12.333713 & 0.4324 $\pm$ 0.0123 $^{\mathrm{p}}$ & 5.2070 $\pm$ 0.3703 & 12.6458 $\pm$ 0.0639 & 11.5854 $\pm$ 0.1110 & 10.8784 $\pm$ 0.0313 \\
J0251$-$2358 $^{\mathrm{o}}$ & 42.990479 & -23.978480 & 0.3488 $\pm$ 0.0075 $^{\mathrm{p}}$ & 11.5809 $\pm$ 0.3037 & 13.2463 $\pm$ 0.0266 & 12.0131 $\pm$ 0.0789 & 11.2374 $\pm$ 0.0235 \\
J0255$-$1640 $^{\mathrm{m}}$ & 43.821469 & -16.681517 & 0.6949 $\pm$ 0.0251 $^{\mathrm{p}}$ & 3.5502 $\pm$ 0.2865 & 12.5358 $\pm$ 0.0759 & 11.5250 $\pm$ 0.1039 & 10.8247 $\pm$ 0.0410 \\
J0257$-$2008 $^{\mathrm{m}}$ & 44.433800 & -20.147361 & 0.7033 $\pm$ 0.0214 $^{\mathrm{p}}$ & 5.5994 $\pm$ 0.2795 & 12.9349 $\pm$ 0.0536 & 11.8526 $\pm$ 0.2819 & 11.2488 $\pm$ 0.0580 \\
J0302$-$1004 $^{\mathrm{m}}$ & 45.597783 & -10.076458 & 0.5506 $\pm$ 0.0644 $^{\mathrm{p}}$ & 17.7469 $\pm$ 0.5137 & 13.8198 $\pm$ 0.0644 & 12.2240 $\pm$ 0.1049 & 11.5825 $\pm$ 0.1280 \\
J0303$-$2119 $^{\mathrm{m}}$ & 45.791775 & -21.325857 & 0.4059 $\pm$ 0.0176 $^{\mathrm{p}}$ & 11.3645 $\pm$ 0.3024 & 13.2950 $\pm$ 0.0317 & 11.8008 $\pm$ 0.1510 & 11.1726 $\pm$ 0.0455 \\
J0310$-$1746 $^{\mathrm{o}}_*$ & 47.708730 & -17.774793 & 0.6691 $\pm$ 0.0524 $^{\mathrm{p}}$ & 2.8197 $\pm$ 0.3446 & 12.3228 $\pm$ 0.1154 & 11.3129 $\pm$ 0.2779 & 10.6511 $\pm$ 0.0904 \\
J0318$-$2915 $^{\mathrm{m}}$ & 49.619953 & -29.259311 & 0.5432 $\pm$ 0.0246 $^{\mathrm{p}}$ & 4.8071 $\pm$ 0.2838 & 12.7956 $\pm$ 0.0588 & 11.6400 $\pm$ 0.1526 & 10.9119 $\pm$ 0.0520 \\
J0327$-$1326 $^{\mathrm{m}}_*$ & 51.863245 & -13.439585 & 0.5759 $\pm$ 0.0230 $^{\mathrm{p}}$ & 17.9639 $\pm$ 0.3194 & 13.8495 $\pm$ 0.0308 & 12.3591 $\pm$ 0.0745 & 11.7381 $\pm$ 0.0406 \\
J0328$-$2140 $^{\mathrm{m}}$ & 52.056629 & -21.672094 & 0.5617 $\pm$ 0.0125 $^{\mathrm{p}}$ & 19.9544 $\pm$ 1.3517 & 13.9310 $\pm$ 0.0627 & 12.4833 $\pm$ 0.1495 & 11.8599 $\pm$ 0.0298 \\
J0334$-$1311 $^{\mathrm{o}}_*$ & 53.625432 & -13.186562 & 0.3622 $\pm$ 0.0121 $^{\mathrm{p}}$ & 6.4737 $\pm$ 0.8420 & 12.8809 $\pm$ 0.1129 & 11.7237 $\pm$ 0.1267 & 10.9540 $\pm$ 0.0395 \\
J0340$-$2533 $^{\mathrm{m}}$ & 55.089050 & -25.558420 & 0.6326 $\pm$ 0.0189 $^{\mathrm{p}}$ & 3.6617 $\pm$ 0.4159 & 12.6341 $\pm$ 0.1005 & 11.4695 $\pm$ 0.1713 & 10.7794 $\pm$ 0.0369 \\
J0347$-$2454 $^{\mathrm{o}}$ & 56.935631 & -24.908777 & 0.6390 $\pm$ 0.0350 $^{\mathrm{p}}$ & 2.7251 $\pm$ 0.2930 & 12.3839 $\pm$ 0.0993 & 11.3985 $\pm$ 0.1457 & 10.7782 $\pm$ 0.0595 \\
J0348$-$2145 $^{\mathrm{m}}$ & 57.009695 & -21.750889 & 0.3565 $\pm$ 0.0173 $^{\mathrm{p}}$ & 9.3671 $\pm$ 0.9747 & 13.0757 $\pm$ 0.0894 & 11.9366 $\pm$ 0.1855 & 11.1970 $\pm$ 0.0648 \\
J0349$-$1500 $^{\mathrm{o}}$ & 57.450133 & -15.002336 & 0.3212 $\pm$ 0.0086 $^{\mathrm{p}}$ & 6.8569 $\pm$ 0.4591 & 12.7580 $\pm$ 0.0603 & 11.6385 $\pm$ 0.0784 & 10.8471 $\pm$ 0.0288 \\
J0353$-$1706 $^{\mathrm{o}}$ & 58.442671 & -17.110893 & 0.5256 $\pm$ 0.0904 $^{\mathrm{p}}$ & 2.1072 $\pm$ 0.2850 & 12.0748 $\pm$ 0.1464 & 11.3161 $\pm$ 0.1087 & 10.4393 $\pm$ 0.2151 \\
J0354$-$1609 $^{\mathrm{o}}_*$ & 58.576148 & -16.164544 & 0.5887 $\pm$ 0.0082 $^{\mathrm{p}}$ & 2.5893 $\pm$ 0.3412 & 12.3025 $\pm$ 0.1157 & 11.5340 $\pm$ 0.1279 & 10.8399 $\pm$ 0.0203 \\
J0400$-$1357 $^{\mathrm{m}}$ & 60.241996 & -13.956629 & 0.6457 $\pm$ 0.0335 $^{\mathrm{p}}$ & 11.0958 $\pm$ 0.6838 & 13.4878 $\pm$ 0.0642 & 11.9762 $\pm$ 0.2079 & 11.2928 $\pm$ 0.0615 \\
J0401$-$0951 $^{\mathrm{o}}$ & 60.314572 & -9.860522 & 0.5816 $\pm$ 0.1755 $^{\mathrm{p}}$ & 3.6594 $\pm$ 0.2919 & 12.5968 $\pm$ 0.1684 & 11.6679 $\pm$ 0.0324 & 10.6850 $\pm$ 0.3761 \\
J0411$-$2256 $^{\mathrm{m}}$ & 62.783524 & -22.944009 & 0.4954 $\pm$ 0.0267 $^{\mathrm{p}}$ & 4.7810 $\pm$ 0.3925 & 12.6322 $\pm$ 0.0774 & 11.6593 $\pm$ 0.1530 & 10.9209 $\pm$ 0.0584 \\
J0420$-$0421 $^{\mathrm{o}}$ & 65.242499 & -4.357444 & 0.7245 $\pm$ 0.0421 $^{\mathrm{p}}$ & 3.0770 $\pm$ 0.3338 & 12.5537 $\pm$ 0.1040 & 11.4806 $\pm$ 0.2010 & 10.8489 $\pm$ 0.0704 \\
J0423$-$0121 $^{\mathrm{o}}$ & 65.796719 & -1.351486 & 0.3851 $\pm$ 0.0245 $^{\mathrm{p}}$ & 3.8592 $\pm$ 0.3035 & 12.3357 $\pm$ 0.0754 & 11.4894 $\pm$ 0.1272 & 10.7525 $\pm$ 0.0713 \\
J0424$-$1624 $^{\mathrm{o}}$ & 66.011723 & -16.411331 & 0.6681 $\pm$ 0.0379 $^{\mathrm{p}}$ & 2.8513 $\pm$ 0.3266 & 12.4460 $\pm$ 0.1052 & 11.4546 $\pm$ 0.1650 & 10.7575 $\pm$ 0.0649 \\
J0424$-$3317 $^{\mathrm{o}}_*$ & 66.161195 & -33.294925 & 0.5542 $\pm$ 0.0594 $^{\mathrm{p}}$ & 5.3286 $\pm$ 0.5552 & 12.8966 $\pm$ 0.1025 & 11.9202 $\pm$ 0.1120 & 11.1285 $\pm$ 0.1143 \\
J0429$-$2957 $^{\mathrm{m}}$ & 67.472905 & -29.959746 & 0.6919 $\pm$ 0.0195 $^{\mathrm{p}}$ & 11.6936 $\pm$ 0.4522 & 13.5673 $\pm$ 0.0439 & 12.0490 $\pm$ 0.1094 & 11.4047 $\pm$ 0.0318 \\
J0432$-$2000 $^{\mathrm{o}}$ & 68.102241 & -20.003820 & 0.5642 $\pm$ 0.0164 $^{\mathrm{p}}$ & 4.3580 $\pm$ 0.4779 & 12.6139 $\pm$ 0.0957 & 11.6367 $\pm$ 0.1024 & 10.9582 $\pm$ 0.0327 \\
J0440$-$2658 $^{\mathrm{m}}$ & 70.192758 & -26.975392 & 0.5018 $\pm$ 0.0089 $^{\mathrm{p}}$ & 10.0192 $\pm$ 0.5052 & 13.2797 $\pm$ 0.0471 & 11.9059 $\pm$ 0.0943 & 11.2297 $\pm$ 0.0201 \\
J0447$-$0251 $^{\mathrm{m}}$ & 71.865678 & -2.861248 & 0.4270 $\pm$ 0.0104 $^{\mathrm{p}}$ & 10.6371 $\pm$ 2.8087 & 13.2880 $\pm$ 0.2274 & 11.9280 $\pm$ 0.4853 & 11.2063 $\pm$ 0.1004 \\
J0450$-$3302 $^{\mathrm{m}}$ & 72.501915 & -33.048279 & 0.4990 $\pm$ 0.0158 $^{\mathrm{p}}$ & 6.8058 $\pm$ 0.2874 & 12.9414 $\pm$ 0.0426 & 11.7995 $\pm$ 0.0827 & 11.0905 $\pm$ 0.0351 \\
J0451$+$0006 $^{\mathrm{m}}_*$ & 72.977704 & 0.105035 & 0.4221 $\pm$ 0.0171 $^{\mathrm{p}}$ & 38.2434 $\pm$ 0.5807 & 14.3657 $\pm$ 0.0262 & 12.4690 $\pm$ 0.1348 & 11.8117 $\pm$ 0.0456 \\
J0452$-$3540 $^{\mathrm{o}}$ & 73.123877 & -35.674171 & 0.5958 $\pm$ 0.0123 $^{\mathrm{p}}$ & 2.9821 $\pm$ 0.2757 & 12.3098 $\pm$ 0.0827 & 11.4428 $\pm$ 0.1050 & 10.7498 $\pm$ 0.0280 \\
J0455$-$2530 $^{\mathrm{m}}_*$ & 73.903064 & -25.512814 & 0.3524 $\pm$ 0.0374 $^{\mathrm{p}}$ & 15.9784 $\pm$ 0.3278 & 13.5320 $\pm$ 0.0486 & 11.7085 $\pm$ 0.1650 & 11.0784 $\pm$ 0.1058 \\
J0458$-$2637 $^{\mathrm{m}}$ & 74.673306 & -26.626873 & 0.2677 $\pm$ 0.0075 $^{\mathrm{p}}$ & 11.3707 $\pm$ 0.2704 & 13.1178 $\pm$ 0.0247 & 11.7328 $\pm$ 0.0738 & 10.9570 $\pm$ 0.0304 \\
J0459$-$3043 $^{\mathrm{m}}$ & 74.964252 & -30.723626 & 0.4329 $\pm$ 0.0106 $^{\mathrm{p}}$ & 3.6167 $\pm$ 0.2939 & 12.3302 $\pm$ 0.0723 & 11.4267 $\pm$ 0.1459 & 10.7054 $\pm$ 0.0283 \\
J0513$-$3050 $^{\mathrm{m}}$ & 78.356121 & -30.843339 & 0.3814 $\pm$ 0.0152 $^{\mathrm{p}}$ & 5.4571 $\pm$ 0.4665 & 12.6339 $\pm$ 0.0773 & 11.5483 $\pm$ 0.1300 & 10.8857 $\pm$ 0.0419 \\
J0524$-$2721 $^{\mathrm{m}}$ & 81.098605 & -27.353147 & 0.3081 $\pm$ 0.0086 $^{\mathrm{p}}$ & 6.8668 $\pm$ 0.3381 & 12.7403 $\pm$ 0.0454 & 11.6454 $\pm$ 0.0767 & 10.8292 $\pm$ 0.0304 \\
J0527$-$1858 $^{\mathrm{m}}_*$ & 81.754744 & -18.967375 & 0.4420 $\pm$ 0.0116 $^{\mathrm{p}}$ & 7.3170 $\pm$ 0.8235 & 13.0720 $\pm$ 0.0995 & 11.7414 $\pm$ 0.1227 & 11.0457 $\pm$ 0.0293 \\
J0532$-$3545 $^{\mathrm{m}}$ & 83.077550 & -35.755547 & 0.4924 $\pm$ 0.0313 $^{\mathrm{p}}$ & 5.3842 $\pm$ 0.2534 & 12.7307 $\pm$ 0.0526 & 11.5237 $\pm$ 0.0856 & 10.8117 $\pm$ 0.0655 \\
J0545$-$2635 $^{\mathrm{m}}_*$ & 86.306565 & -26.588383 & 0.3177 $\pm$ 0.0220 $^{\mathrm{p}}$ & 13.3520 $\pm$ 0.3016 & 13.3301 $\pm$ 0.0347 & 11.9454 $\pm$ 0.0871 & 11.3816 $\pm$ 0.0687 \\
J0545$-$3542 $^{\mathrm{o}}$ & 86.352690 & -35.714575 & 0.4065 $\pm$ 0.0086 $^{\mathrm{p}}$ & 10.8868 $\pm$ 0.3253 & 13.2580 $\pm$ 0.0308 & 11.5275 $\pm$ 0.1980 & 10.9691 $\pm$ 0.0232 \\
J0549$-$2355 $^{\mathrm{m}}$ & 87.402875 & -23.925304 & 0.5165 $\pm$ 0.0114 $^{\mathrm{p}}$ & 5.7713 $\pm$ 0.6578 & 12.8185 $\pm$ 0.1016 & 11.8273 $\pm$ 0.2088 & 11.1444 $\pm$ 0.0383 \\
J0553$-$2853 $^{\mathrm{o}}$ & 88.296789 & -28.893375 & 0.7899 $\pm$ 0.0617 $^{\mathrm{p}}$ & 2.4014 $\pm$ 0.2957 & 12.3890 $\pm$ 0.1217 & 11.0999 $\pm$ 0.1918 & 10.6506 $\pm$ 0.0986 \\
J0559$-$3540 $^{\mathrm{o}}_*$ & 89.956356 & -35.670137 & 0.8034 $\pm$ 0.0381 $^{\mathrm{p}}$ & 2.5171 $\pm$ 0.3366 & 12.4390 $\pm$ 0.1229 & 11.5070 $\pm$ 0.2653 & 10.9215 $\pm$ 0.0579 \\
J1132$+$0212 $^{\mathrm{o}}_*$ & 173.111909 & 2.215971 & 0.9367 $\pm$ 0.0505 $^{\mathrm{p}}$ & 11.7490 $\pm$ 0.8730 & 13.7452 $\pm$ 0.0822 & 11.7838 $\pm$ 0.3088 & 11.3107 $\pm$ 0.0719 \\
J2043$-$0609 $^{\mathrm{o}}_*$ & 310.801950 & -6.164955 & 0.8125 $\pm$ 0.0385 $^{\mathrm{p}}$ & 4.9476 $\pm$ 0.4128 & 12.9076 $\pm$ 0.0820 & 11.7554 $\pm$ 0.1246 & 11.1610 $\pm$ 0.0518 \\
J2105$+$0537 $^{\mathrm{m}}$ & 316.420521 & 5.620318 & 0.3095 $\pm$ 0.0179 $^{\mathrm{p}}$ & 5.5730 $\pm$ 0.3614 & 12.5595 $\pm$ 0.0626 & 11.8111 $\pm$ 0.1278 & 11.0619 $\pm$ 0.0645 \\
J2106$-$0027 $^{\mathrm{m}}$ & 316.613480 & -0.458322 & 0.7903 $\pm$ 0.0313 $^{\mathrm{p}}$ & 7.2352 $\pm$ 0.4622 & 13.2220 $\pm$ 0.0650 & 11.8819 $\pm$ 0.2064 & 11.3414 $\pm$ 0.0494 \\
J2106$-$0547 $^{\mathrm{m}}$ & 316.721396 & -5.783672 & 0.6485 $\pm$ 0.0501 $^{\mathrm{p}}$ & 3.9670 $\pm$ 0.9756 & 12.7354 $\pm$ 0.2151 & 11.5964 $\pm$ 0.2627 & 10.9210 $\pm$ 0.1069 \\
J2109$-$0135 $^{\mathrm{o}}$ & 317.329931 & -1.593868 & 0.8934 $\pm$ 0.0533 $^{\mathrm{p}}$ & 1.6177 $\pm$ 0.3466 & 12.0116 $\pm$ 0.1874 & 11.3098 $\pm$ 0.4148 & 10.6771 $\pm$ 0.1094 \\
J2111$-$0114 $^{\mathrm{m}}_*$ & 317.830618 & -1.239845 & 0.6386 $\pm$ 0.0001 $^{\mathrm{s}}$ & 12.6015 $\pm$ 1.5271 & 13.5989 $\pm$ 0.1090 & 12.1983 $\pm$ 0.1970 & 11.5051 $\pm$ 0.0348 \\
J2114$+$0658 $^{\mathrm{m}}$ & 318.724767 & 6.968209 & 0.2081 $\pm$ 0.0052 $^{\mathrm{p}}$ & 8.1920 $\pm$ 0.4988 & 12.7287 $\pm$ 0.0552 & 11.5572 $\pm$ 0.0729 & 10.7607 $\pm$ 0.0261 \\
J2126$+$0949 $^{\mathrm{o}}$ & 321.560372 & 9.833233 & 0.8025 $\pm$ 0.0291 $^{\mathrm{p}}$ & 4.2152 $\pm$ 0.5613 & 12.8836 $\pm$ 0.1217 & 11.5651 $\pm$ 0.2346 & 11.0047 $\pm$ 0.0434 \\
J2129$-$0126 $^{\mathrm{m}}$ & 322.287133 & -1.442114 & 0.9581 $\pm$ 0.0949 $^{\mathrm{p}}$ & 17.5135 $\pm$ 0.2840 & 14.1093 $\pm$ 0.0862 & 11.8993 $\pm$ 0.3343 & 11.3685 $\pm$ 0.1279 \\
J2143$+$1431 $^{\mathrm{m}}$ & 325.794687 & 14.522257 & 0.4102 $\pm$ 0.0087 $^{\mathrm{p}}$ & 9.4958 $\pm$ 0.2964 & 13.1429 $\pm$ 0.0315 & 11.5898 $\pm$ 0.2050 & 11.0386 $\pm$ 0.0231 \\
J2146$-$0317 $^{\mathrm{m}}$ & 326.623059 & -3.284784 & 0.5759 $\pm$ 0.0213 $^{\mathrm{p}}$ & 12.3891 $\pm$ 0.2969 & 13.5277 $\pm$ 0.0324 & 11.8411 $\pm$ 0.0991 & 11.2853 $\pm$ 0.0397 \\
J2151$-$0138 $^{\mathrm{m}}$ & 327.858149 & -1.647066 & 0.3131 $\pm$ 0.0001 $^{\mathrm{s}}$ & 15.8904 $\pm$ 0.2973 & 13.4747 $\pm$ 0.0183 & 12.1191 $\pm$ 0.0941 & 11.3451 $\pm$ 0.0036 \\
J2206$+$1104 $^{\mathrm{o}}$ & 331.567919 & 11.068702 & 0.7772 $\pm$ 0.0002 $^{\mathrm{s}}$ & 2.9128 $\pm$ 0.2950 & 12.4242 $\pm$ 0.0923 & 11.8076 $\pm$ 0.1569 & 11.0927 $\pm$ 0.0235 \\
J2207$-$0411 $^{\mathrm{o}}_*$ & 331.762532 & -4.196620 & 0.7576 $\pm$ 0.0416 $^{\mathrm{p}}$ & 3.6141 $\pm$ 0.4805 & 12.7173 $\pm$ 0.1233 & 11.4490 $\pm$ 0.2719 & 10.9727 $\pm$ 0.0641 \\
J2210$+$2604 $^{\mathrm{m}}$ & 332.512025 & 26.068904 & 0.3467 $\pm$ 0.0001 $^{\mathrm{s}}$ & 7.4952 $\pm$ 0.3672 & 12.8665 $\pm$ 0.0435 & 11.7216 $\pm$ 0.0811 & 10.8841 $\pm$ 0.0049 \\
J2222$+$2745 $^{\mathrm{m}}$ & 335.536310 & 27.759241 & 0.4717 $\pm$ 0.0404 $^{\mathrm{p}}$ & 7.4156 $\pm$ 0.7214 & 12.9935 $\pm$ 0.0944 & 12.0032 $\pm$ 0.1067 & 11.2539 $\pm$ 0.0986 \\
J2226$+$0041 $^{\mathrm{m}}$ & 336.538789 & 0.694997 & 0.6471 $\pm$ 0.0001 $^{\mathrm{s}}$ & 3.1648 $\pm$ 0.2902 & 12.4011 $\pm$ 0.0821 & 11.8315 $\pm$ 0.0830 & 10.9637 $\pm$ 0.0072 \\
J2234$-$0630 $^{\mathrm{o}}_*$ & 338.724911 & -6.515025 & 0.8146 $\pm$ 0.0705 $^{\mathrm{p}}$ & 2.5401 $\pm$ 0.3858 & 12.4585 $\pm$ 0.1475 & 11.3684 $\pm$ 0.3503 & 10.7567 $\pm$ 0.1033 \\
J2238$+$1319 $^{\mathrm{m}}_*$ & 339.630481 & 13.332188 & 0.4129 $\pm$ 0.0001 $^{\mathrm{s}}$ & 9.3379 $\pm$ 0.2844 & 13.2490 $\pm$ 0.0289 & 11.8872 $\pm$ 0.2588 & 11.3220 $\pm$ 0.0034 \\
J2240$-$0139 $^{\mathrm{o}}$ & 340.054725 & -1.659645 & 0.8515 $\pm$ 0.0393 $^{\mathrm{p}}$ & 8.4577 $\pm$ 0.6090 & 13.4011 $\pm$ 0.0781 & 11.5669 $\pm$ 0.1932 & 11.0167 $\pm$ 0.0603 \\
J2241$-$0527 $^{\mathrm{m}}$ & 340.298713 & -5.460154 & 0.9632 $\pm$ 0.0418 $^{\mathrm{p}}$ & 8.2528 $\pm$ 0.3293 & 13.4576 $\pm$ 0.0675 & 12.1399 $\pm$ 0.3072 & 11.4125 $\pm$ 0.0614 \\
J2244$+$2759 $^{\mathrm{o}}$ & 341.020635 & 27.987671 & 0.3429 $\pm$ 0.0001 $^{\mathrm{s}}$ & 6.1671 $\pm$ 0.3529 & 12.6923 $\pm$ 0.0502 & 11.7762 $\pm$ 0.0629 & 10.9361 $\pm$ 0.0075 \\
J2246$+$0415 $^{\mathrm{o}}$ & 341.687109 & 4.263947 & 1.0241 $\pm$ 0.0503 $^{\mathrm{p}}$ & 9.7144 $\pm$ 0.4445 & 13.6397 $\pm$ 0.0822 & 11.7668 $\pm$ 0.3664 & 11.3853 $\pm$ 0.0666 \\
J2247$-$0205 $^{\mathrm{m}}$ & 341.801229 & -2.093837 & 0.3383 $\pm$ 0.0234 $^{\mathrm{p}}$ & 8.7637 $\pm$ 0.3311 & 12.9922 $\pm$ 0.0454 & 11.8279 $\pm$ 0.1308 & 11.1118 $\pm$ 0.0712 \\
J2248$-$0123 $^{\mathrm{m}}$ & 342.151533 & -1.392814 & 0.3983 $\pm$ 0.0001 $^{\mathrm{s}}$ & 4.1811 $\pm$ 0.3079 & 12.4201 $\pm$ 0.0642 & 11.4759 $\pm$ 0.1523 & 10.6602 $\pm$ 0.0064 \\
J2248$+$2015 $^{\mathrm{o}}$ & 342.151611 & 20.252945 & 0.8046 $\pm$ 0.0320 $^{\mathrm{p}}$ & 4.0677 $\pm$ 0.2971 & 12.7327 $\pm$ 0.0747 & 11.5915 $\pm$ 0.1122 & 11.0353 $\pm$ 0.0461 \\
J2249$-$0110 $^{\mathrm{o}}$ & 342.410083 & -1.179021 & 0.3399 $\pm$ 0.0001 $^{\mathrm{s}}$ & 4.3863 $\pm$ 0.3032 & 12.3940 $\pm$ 0.0610 & 11.3715 $\pm$ 0.0718 & 10.5252 $\pm$ 0.0072 \\
J2252$-$0413 $^{\mathrm{m}}$ & 343.040271 & -4.218693 & 0.4273 $\pm$ 0.0139 $^{\mathrm{p}}$ & 7.5307 $\pm$ 0.3258 & 12.9609 $\pm$ 0.0417 & 11.7440 $\pm$ 0.1254 & 11.0997 $\pm$ 0.0368 \\
J2256$+$1005 $^{\mathrm{m}}$ & 344.205409 & 10.086293 & 0.8824 $\pm$ 0.0395 $^{\mathrm{p}}$ & 7.2892 $\pm$ 0.5472 & 13.2937 $\pm$ 0.0791 & 11.7391 $\pm$ 0.1570 & 11.2694 $\pm$ 0.0516 \\
J2257$-$0610 $^{\mathrm{o}}_*$ & 344.434506 & -6.173864 & 0.7776 $\pm$ 0.0335 $^{\mathrm{p}}$ & 5.1914 $\pm$ 0.3931 & 12.9247 $\pm$ 0.0780 & 11.5515 $\pm$ 0.2797 & 10.9299 $\pm$ 0.0522 \\
J2258$+$1709 $^{\mathrm{m}}$ & 344.556726 & 17.150617 & 0.7990 $\pm$ 0.0413 $^{\mathrm{p}}$ & 8.5651 $\pm$ 0.3165 & 13.3727 $\pm$ 0.0560 & 11.9614 $\pm$ 0.2630 & 11.4640 $\pm$ 0.0667 \\
J2259$+$1212 $^{\mathrm{m}}$ & 344.797839 & 12.209790 & 0.8885 $\pm$ 0.1309 $^{\mathrm{p}}$ & 4.8344 $\pm$ 0.4936 & 13.0611 $\pm$ 0.1382 & 11.8307 $\pm$ 0.2041 & 11.0082 $\pm$ 0.1775 \\
J2300$+$2213 $^{\mathrm{m}}$ & 345.071926 & 22.224908 & 0.4237 $\pm$ 0.0116 $^{\mathrm{p}}$ & 17.8792 $\pm$ 0.3283 & 13.7075 $\pm$ 0.0232 & 12.1271 $\pm$ 0.1368 & 11.4803 $\pm$ 0.0339 \\
J2303$+$2328 $^{\mathrm{m}}$ & 345.860959 & 23.475984 & 0.2767 $\pm$ 0.0000 $^{\mathrm{s}}$ & 9.0895 $\pm$ 0.4214 & 12.9385 $\pm$ 0.0413 & 11.6932 $\pm$ 0.0904 & 10.9166 $\pm$ 0.0066 \\
J2304$+$3327 $^{\mathrm{o}}_*$ & 346.022626 & 33.459947 & 0.4837 $\pm$ 0.0233 $^{\mathrm{p}}$ & 7.5359 $\pm$ 0.9981 & 13.0214 $\pm$ 0.1149 & 11.7940 $\pm$ 0.1222 & 11.0639 $\pm$ 0.0527 \\
J2304$-$0052 $^{\mathrm{m}}$ & 346.101457 & -0.877417 & 0.8850 $\pm$ 0.0305 $^{\mathrm{p}}$ & 4.9139 $\pm$ 0.3447 & 12.9512 $\pm$ 0.0742 & 11.7667 $\pm$ 0.0970 & 11.1624 $\pm$ 0.0436 \\
J2307$-$1322 $^{\mathrm{m}}_*$ & 346.766785 & -13.374414 & 0.5087 $\pm$ 0.0223 $^{\mathrm{p}}$ & 7.6091 $\pm$ 0.3016 & 13.0470 $\pm$ 0.0429 & 11.8757 $\pm$ 0.0725 & 11.1635 $\pm$ 0.0504 \\
J2308$-$0211 $^{\mathrm{m}}$ & 347.092613 & -2.192172 & 0.2949 $\pm$ 0.0097 $^{\mathrm{p}}$ & 37.4668 $\pm$ 1.5427 & 14.1950 $\pm$ 0.0394 & 12.3581 $\pm$ 0.1176 & 11.6438 $\pm$ 0.0358 \\
J2312$+$0451 $^{\mathrm{m}}$ & 348.245242 & 4.861479 & 0.3102 $\pm$ 0.0001 $^{\mathrm{s}}$ & 9.5857 $\pm$ 0.3136 & 13.0322 $\pm$ 0.0298 & 11.7470 $\pm$ 0.1007 & 10.9685 $\pm$ 0.0099 \\
J2313$-$0104 $^{\mathrm{m}}$ & 348.477049 & -1.080103 & 0.5312 $\pm$ 0.0002 $^{\mathrm{s}}$ & 7.8778 $\pm$ 0.3529 & 13.0961 $\pm$ 0.0423 & 11.8164 $\pm$ 0.0840 & 11.1070 $\pm$ 0.0069 \\
J2318$-$1106 $^{\mathrm{o}}$ & 349.549151 & -11.101265 & 0.7203 $\pm$ 0.0264 $^{\mathrm{p}}$ & 5.4424 $\pm$ 0.2841 & 12.9250 $\pm$ 0.0550 & 11.8356 $\pm$ 0.1223 & 11.1832 $\pm$ 0.0401 \\
J2319$+$0038 $^{\mathrm{m}}$ & 349.972629 & 0.637053 & 0.8295 $\pm$ 0.1561 $^{\mathrm{p}}$ & 8.9365 $\pm$ 1.5724 & 13.4602 $\pm$ 0.1998 & 12.0127 $\pm$ 0.4047 & 11.4754 $\pm$ 0.2369 \\
J2320$-$1202 $^{\mathrm{m}}$ & 350.027080 & -12.034294 & 0.3967 $\pm$ 0.0130 $^{\mathrm{p}}$ & 5.6349 $\pm$ 0.3893 & 12.6764 $\pm$ 0.0618 & 11.6129 $\pm$ 0.1852 & 10.8974 $\pm$ 0.0531 \\
J2326$+$2026 $^{\mathrm{m}}$ & 351.748351 & 20.449553 & 0.7895 $\pm$ 0.0289 $^{\mathrm{p}}$ & 5.7885 $\pm$ 0.5071 & 13.0265 $\pm$ 0.0839 & 11.8550 $\pm$ 0.1355 & 11.1777 $\pm$ 0.0465 \\
J2331$+$2749 $^{\mathrm{m}}$ & 352.825967 & 27.817418 & 0.8688 $\pm$ 0.0618 $^{\mathrm{p}}$ & 9.5300 $\pm$ 0.3054 & 13.5149 $\pm$ 0.0627 & 11.5676 $\pm$ 0.2760 & 11.2313 $\pm$ 0.0854 \\
J2334$-$0746 $^{\mathrm{o}}$ & 353.528695 & -7.771200 & 0.4018 $\pm$ 0.0130 $^{\mathrm{p}}$ & 5.6608 $\pm$ 0.4067 & 12.6863 $\pm$ 0.0656 & 11.4975 $\pm$ 0.1452 & 10.7977 $\pm$ 0.0334 \\
J2335$+$0922 $^{\mathrm{o}}$ & 353.981202 & 9.382482 & 0.7536 $\pm$ 0.0302 $^{\mathrm{p}}$ & 6.6283 $\pm$ 0.2919 & 13.1196 $\pm$ 0.0537 & 11.4743 $\pm$ 0.1691 & 10.9540 $\pm$ 0.0480 \\
J2347$-$0047 $^{\mathrm{m}}$ & 356.811434 & -0.797979 & 0.7889 $\pm$ 0.0188 $^{\mathrm{p}}$ & 8.1732 $\pm$ 0.3176 & 13.3238 $\pm$ 0.0493 & 11.9933 $\pm$ 0.1101 & 11.3661 $\pm$ 0.0289 \\
J2347$-$0439 $^{\mathrm{o}}$ & 356.954863 & -4.650060 & 0.9785 $\pm$ 0.0842 $^{\mathrm{p}}$ & 2.7527 $\pm$ 0.2918 & 12.5209 $\pm$ 0.1247 & 11.3972 $\pm$ 0.3590 & 10.8321 $\pm$ 0.1066 \\
J2348$+$1407 $^{\mathrm{o}}$ & 357.034778 & 14.129527 & 0.6546 $\pm$ 0.0002 $^{\mathrm{s}}$ & 2.5010 $\pm$ 0.4193 & 12.3271 $\pm$ 0.1451 & 11.5617 $\pm$ 0.1332 & 10.7968 $\pm$ 0.0200 \\
J2356$+$0241 $^{\mathrm{o}}$ & 359.142118 & 2.692205 & 0.8204 $\pm$ 0.0304 $^{\mathrm{p}}$ & 4.9951 $\pm$ 1.1665 & 13.0621 $\pm$ 0.2101 & 11.3851 $\pm$ 0.3041 & 10.8635 $\pm$ 0.0601 \\
J2359$-$1214 $^{\mathrm{m}}$ & 359.802678 & -12.236703 & 0.9131 $\pm$ 0.0672 $^{\mathrm{p}}$ & 7.4873 $\pm$ 0.3129 & 13.3384 $\pm$ 0.0778 & 11.6796 $\pm$ 0.3689 & 11.2509 $\pm$ 0.0906 \\
J2359$+$0208 $^{\mathrm{m}}_*$ & 359.889762 & 2.139947 & 0.4294 $\pm$ 0.0001 $^{\mathrm{s}}$ & 9.8667 $\pm$ 0.7730 & 13.1989 $\pm$ 0.0693 & 12.1816 $\pm$ 0.1346 & 11.4537 $\pm$ 0.0109 \\
\enddata
\tablecomments{All values quoted in dex are constrained exclusively within the Einstein aperture for each system. $M_{\odot,\Sigma}$ and $M_{\odot,\star}$ refer to total mass and stellar mass, respectively. Luminosity was measured within the rest-frame wavelength interval of 3000--7000\r{A}. Superscript labels in the ID column are as follows: `o' and `m' stand for whether one or multiple cluster members fall within the Einstein aperture, respectively. Subscript asterisks in the ID column denote systems that were analyzed with \texttt{LENSTOOL} in Section \ref{lenstool}. Superscript labels in the $z_{_{BCG}}$ column are as follows: `p' stands for LS DR9 photometric redshift, and `s' stands for SDSS DR15 spectroscopic redshift. Table \ref{tab:sample} is published in machine-readable format. (The machine-readable version of this Table is available in the \href{https://doi.org/10.3847/1538-4357/ada24c}{online article}.)}
\end{deluxetable*}

\allauthors
\end{document}